\documentclass[12pt,bibliography]{article}
\usepackage{hyperref}
\usepackage{epsfig}
\usepackage{amsfonts}
\usepackage{amsmath}

\newcommand{\be}{\begin{equation}}
\newcommand{\ee}{\end{equation}}
\newcommand{\ba}{\begin{eqnarray}}
\newcommand{\ea}{\end{eqnarray}}
\newcommand{\bi}{\begin{itemize}}
\newcommand{\ei}{\end{itemize}}

\newcommand{\Ncal}{{\mathcal N}}
\newcommand{\Ocal}{{\mathcal O}}

\newcommand{\aslash}[1]{\,\,{\raise.15ex\hbox{/}\mkern-12mu #1}}
\newcommand{\bslash}[1]{\,\,{\raise.15ex\hbox{/}\mkern-9mu #1}}

\renewcommand{\bar}{\overline}

\renewcommand{\hat}{\widehat}
\renewcommand{\Im}{{\rm Im\,}}
\renewcommand{\Re}{{\rm Re\,}}

\newcommand\lrpar{\raise .8ex\hbox{$^\leftrightarrow$} \hspace{-9pt}
\partial}
\newcommand\lpar{\raise .8ex\hbox{$^\leftarrow$} \hspace{-9pt}
\partial}
\newcommand\rpar{\raise .8ex\hbox{$^\rightarrow$} \hspace{-9pt}
\partial}
\newcommand\lrd{\raise .8ex\hbox{$^\leftrightarrow$} \hspace{-9pt}
\nabla}

\newcommand{\gsim}{\lower.7ex\hbox{$\;\stackrel{\textstyle>}{\sim}\;$}}
\newcommand{\lsim}{\lower.7ex\hbox{$\;\stackrel{\textstyle<}{\sim}\;$}}
\begin{document}

\baselineskip=18pt

\setcounter{footnote}{0}
\setcounter{figure}{0}
\setcounter{table}{0}

\begin{titlepage}

\begin{center}
\vspace{1cm}

{\Large \bf  Writing CFT correlation functions as AdS scattering amplitudes}

\vspace{0.8cm}

{\bf Joao Penedones}

\vspace{.5cm}

{\it Perimeter Institute for Theoretical Physics, 
Waterloo, Ontario
N2L 2Y5, Canada \\
Kavli Institute for Theoretical Physics,
  Santa Barbara, CA 93106-4030, USA\\
Centro de F\'isica do Porto,
Rua do Campo Alegre 687, 4169-007 Porto, Portugal
}

\end{center}
\vspace{1cm}

\begin{abstract}
We explore the Mellin representation of conformal correlation functions
recently proposed by Mack. Examples in the AdS/CFT context reinforce
the analogy between Mellin amplitudes and scattering amplitudes.
We conjecture a simple formula relating the bulk scattering amplitudes to the asymptotic
behavior of Mellin amplitudes and show that 
previous results on the flat space limit of AdS
follow from our new formula.
We find that the Mellin amplitudes are particularly useful
in the case of conformal gauge theories in the planar limit.
In this case, the four point Mellin amplitudes are meromorphic functions
whose poles and their residues are entirely determined by two and three
point functions of single-trace operators. 
This makes the Mellin amplitudes the ideal objects to attempt
the conformal bootstrap program in higher dimensions.
 
\end{abstract}

\bigskip
\bigskip


\end{titlepage}

\vfill\eject

\section{Introduction}

Scattering amplitudes are transition amplitudes between states that
describe non-interacting and uncorrelated particles in the infinite past ({\it in} states) and
states that
describe non-interacting and uncorrelated particles in the infinite future ({\it out} states).
This definition makes sense in Minkowski spacetime, where particles become infinitely distant from each other in the infinite past and future.
Anti-de Sitter (AdS) spacetime has a timelike conformal boundary and does not admit {\it in} and {\it out}  states.
Pictorically, one can say that particles in AdS live in an box and interact forever.
Thus, in AdS, we can not use the standard definition of scattering amplitudes.
However, we can create and anihilate particles in AdS by changing the boundary conditions at the timelike boundary.
By the AdS/CFT correspondence \cite{Maldacena,GKP,Witten}, the transition amplitudes between this type of states are equal
to the correlation functions of the dual conformal field theory (CFT).
This suggests that we should interpret the CFT correlation functions as AdS scattering amplitudes \cite{Polchinski,Susskind,GiddingsBSM,GiddingsBulkloc}.
In this paper, we support this view using a representation of the conformal correlation functions that makes their scattering amplitude
nature more transparent.

We shall use the Mellin representation  
recently proposed by Mack in \cite{Mack,MackSummary}. 
\footnote{The Mellin representation was used before, for example in \cite{Symanzik,Sofia,Liu},
but its analogy with scattering amplitudes was not emphasized.}
The Euclidean correlator of primary scalar operators
\be
A(x_i)=\langle \Ocal_1(x_1) \dots \Ocal_n(x_n) \rangle\ ,
\ee
can be written as
\begin{align}
A(x_i)=\frac{\Ncal }{(2\pi i)^{n(n-3)/2}} \int  d \delta_{ij}\,
M(\delta_{ij})\,
\prod_{i<j}^n \Gamma(\delta_{ij}) \left( x_{ij}^2 \right)^{-\delta_{ij}} \label{genMellin}
\end{align}
where the integration contour runs parallel to the imaginary axis with 
$\Re \delta_{ij}>0$.
Moreover, the integration variables are constrained by
\begin{align}
\sum_{j\neq i}^n \delta_{ij}= \Delta_i\ , \label{constraint}
\end{align}
so that  the integrand is conformally covariant with scaling dimension $\Delta_i$ at the point $x_i$.
This gives $n(n-3)/2$ independent integration variables.
We give the precise definition of the integration measure in appendix \ref{appmeasure}.
Notice that $n(n-3)/2$ is also the number of independent conformal invariant cross-ratios that one can make using $n$ points
and the number of independent Mandelstam invariants of a $n$-particle scattering process.
The normalization constant $\Ncal$ will be fixed in the next section.
It is instructive to solve the constraints (\ref{constraint}) using $n$ Lorentzian vectors $k_i$
subject to $\sum_{i=1}^n k_i=0$ and $-k_i^2=\Delta_i$.
Then
\be
\delta_{ij}=k_i\cdot k_j=\frac{\Delta_i+\Delta_j-s_{ij}}{2}\ ,   \label{deltatos}
\ee
with $s_{ij}=-(k_i+k_j)^2$, automatically solves the constraints (\ref{constraint}).

Mack realized that there is a strong similarity between the Mellin amplitude $M(s_{ij})$ and  $n$-particle flat space
scattering amplitudes as functions of the Mandelstam invariants.
In particular, by studying the Mellin representation (\ref{genMellin}) of the  
conformal partial wave
decomposition of the four point function, 
Mack showed that $M(s_{ij})$ is crossing symmetric and meromorphic with simple poles at
\be
s_{13}= \Delta_k-l_k +2m\ , \ \ \ \ \ \ \ \ \ \ m=0,1,2,\dots \ .\label{polesinMellin}
\ee
Here, $\Delta_k$ and $l_k$ are the scaling dimension and spin
 of an operator $\Ocal_k$ present in the
operator product expansions $\Ocal_1\Ocal_3\sim C_{13k}\Ocal_k$ and $\Ocal_2\Ocal_4\sim C_{24k}\Ocal_k$.
Moreover, the residue of the leading pole ($m=0$) is given by the product of the two three point couplings $C_{13k}C_{24k}$ times a known polynomial of degree $l_k$ in the variable $\gamma_{13}=(s_{12}-s_{14})/2$. 
The satellite poles ($m>0$) are determined by the leading one.
In other words, the Mellin amplitude $M(s_{ij})$ obeys exact duality.
 
In this paper we propose that  the Mellin amplitude $M(s_{ij})$ should be taken as the AdS scattering amplitude.
We motivate this proposal with two observations.
Firstly, we compute the Mellin amplitudes $M(s_{ij})$ 
for several Witten diagrams and obtain 
expressions resembling scattering amplitudes in flat space.
For example, we find that contact interactions give rise to polynomial
Mellin amplitudes in perfect analogy with flat space scattering amplitudes
(section \ref{AdSexamples}).
 Notice that the OPE analysis of 
these Witten diagrams contains primary double-trace
 operators  with spin $l$ and conformal dimension
\be
\Delta_i+\Delta_j+l+2p +O(1/N^2) \ ,\ \ \ \ \ \ \ \ p=0,1,2,\dots\ ,
\ee
where $1/N^2$ denotes the coupling constant in AdS \cite{Liu, ourCPW, JP}.
Interestingly, these do not give rise to poles in  $M(s_{ij})$. 
From (\ref{polesinMellin}), at large $N$, one would expect poles at
$
\delta_{ij} = 0,-1,-2,\dots\ ,
$ 
but these are already produced by the $\Gamma$-functions in (\ref{genMellin}).
This suggests that the Mellin representation is particularly useful for CFT's with a weakly coupled bulk dual.
\footnote{Usually this corresponds to a large-N expansion of the CFT.}
We also compute Mellin amplitudes associated with   tree level exchange diagrams
in AdS and verify that all poles are associated to  single-trace operators dual to  
fields exchanged in AdS. 
In section \ref{sectiononeloop}, we determine the Mellin amplitude of a one-loop diagram in 
AdS. In this case, we find that the two particle state exchanged in the loop gives rise to poles of the Mellin amplitude. These examples suggest that we should think of the Mellin amplitude as an amputated amplitude.

A particular example, that illustrates the remarkable simplicity of the Mellin
amplitudes is the graviton exchange between minimally coupled massless  scalars
in AdS$_5$ ($\Delta_i=d=4$).
This Witten diagram was computed in \cite{D'Hoker} in terms of D-functions,
\ba
A(x_i)&\propto& 
9 D_{4 4 4 4}(x_i) -\frac{4 }{3 x_{1 3}^6}D_{1 4 1 4}(x_i)-\frac{20 }{9 x_{1 3}^4}D_{2 4 2 4}(x_i) -\frac{23 }{9 x_{1 3}^2}D_{3 4 3 4}(x_i) 
 \nonumber\\&&
  +\frac{16  (x_{1 4}^2 x_{2 3}^2+  x_{1 2}^2 x_{3 4}^2 )}{3 x_{1 3}^6}D_{2 5 2 5}(x_i)
   +\frac{64 ( x_{1 4}^2 x_{2 3}^2+  x_{1 2}^2 x_{3 4}^2 )}{9 x_{1 3}^4}D_{3 5 3 5}(x_i) 
   \label{Agraviton}
   \\&&
  +\frac{8  ( x_{1 4}^2   x_{2 3}^2+ x_{1 2}^2 x_{3 4}^2- x_{2 4}^2x_{1 3}^2   )}{x_{1 3}^2}D_{4 5 4 5}(x_i)
  \ .
    \nonumber
\ea
We shall give the precise definition of the  D-functions 
in the next section, but
for now it is enough to know that they are given by a non-trivial integral 
representation. 
The result (\ref{Agraviton}) looks quite cumbersome but
the associated Mellin amplitude is  a simple rational function,
\be
M(s_{ij}) \propto
\frac{6 \gamma_{13}^2+2}{s_{13}-2}+
\frac{8 \gamma_{13}^2}{s_{13}-4}+
\frac{\gamma_{13}^2-1}{s_{13}-6}
-\frac{15}{4} s_{13}+\frac{55}{2}
  \ . \label{gravitonMellin}
\ee
This function only has poles at $s_{13}=2,4,6$ contrary to the general 
expectation (\ref{polesinMellin}) of an infinite series of poles at $s_{13}=2+2m$ with $m=0,1,2,\dots$,
associated with the energy-momentum tensor.
In this particular case, there is an extra simplification
and the residues vanish for $m\ge 3$. Furthermore, notice that the residues of the poles are quadratic polynomials in $\gamma_{13}$ as predicted by Mack for spin 2 exchanges.

Secondly, we conjecture that the bulk flat space scattering amplitude $T$ is
encoded in the large $s_{ij}$ limit of the Mellin amplitude $M(s_{ij})$ by the simple
formula
\be
M(s_{ij})\approx \frac{R^{n(1-d)/2+d+1}}{\Gamma \left(\frac{1}{2}\sum_i \Delta_i -\frac{d}{2}\right)}
\int\limits_0^\infty  d\beta\,  \beta^{\frac{1}{2}\sum_i \Delta_i -\frac{d}{2}-1}e^{-\beta} \,T\left(S_{ij}= \frac{2\beta }{R^2}   s_{ij}\right)\ ,\ \ \ \ \ \ \   s_{ij} \gg 1\ ,
\label{mainformula}
\ee
where $S_{ij}=-(K_i+K_j)^2$ are the Mandelstam invariants of the flat space scattering process and $R$ is the AdS radius.
This formula assumes that all external particles become massless under the flat space limit.
In section \ref{FSlimit}, we check that this conjecture is consistent with previous
 studies  \cite{Polchinski, Susskind, GGP, TakuyaFSL,Katz} of the flat space limit of AdS/CFT. 
 In particular, we rederive the results of \cite{TakuyaFSL} starting from (\ref{mainformula}).
 We conclude in section \ref{conclusion} by discussing possible future applications  of the Mellin representation of CFT correlation functions.

\section{Mellin representation of Witten diagrams}
\label{AdSexamples}
We shall start by computing the Mellin representation (\ref{genMellin}) of some simple Witten diagrams.
This will illustrate the simplicity of this representation and
give us enough intuition to help us guess its relation to the flat space limit.

The computation of Witten diagrams is significantly simplified by the use of the embedding space formalism \cite{Dirac, Sofia, Mythesis},
which we quickly review.
Let us consider Euclidean AdS$_{d+1}$ defined by the hyperboloid
\be
X^2=-R^2\ ,\ \ \ \ \ \ \ \ \ \  X^0 > 0\ ,\ \ \ \ \ \ \ \ \ \ \ X \in \mathbb{M}^{d+2}\ ,
\ee
embedded in $(d+2)$-dimensional Minkowski spacetime.
It is convenient to think of the conformal boundary of AdS as the space of null rays
\be
P^2=0\ ,\ \ \ \ \ \ \ \ \ \  P\sim \lambda P\ \ (\lambda \in \mathbb{R})\ ,\ \ \ \ \ \ \ \ \ \ \ P \in \mathbb{M}^{d+2}\ .
\ee
Then, the correlations functions of the dual CFT are encoded into $SO(1,d+1)$ invariant functions of the external points $P_i$, transforming
homogeneously with weights $\Delta_i$.
To recover the usual expressions in physical  $\mathbb{R}^d$  we choose the light cone section
\be
P=(P^+,P^-,P^\mu)=(1,x^2,x^\mu)\ ,
\ee
where $\mu=0,1,\dots,d-1$. This gives
\be
P_{ij}=-2P_i \cdot P_j =(x_i-x_j)^2\ .
\ee


The basic ingredient required to compute Witten diagrams is the bulk to boundary propagator which in this notation is simply given by
\be
G_{B\partial}(X,P)=\frac{\mathcal{C}_\Delta }{R^{(d-1)/2}
(-2 P \cdot X/R)^{\Delta}} =
\frac{\mathcal{C}_\Delta}{R^{(d-1)/2}\Gamma(\Delta)} \int_0^\infty \frac{dt}{t} t^\Delta e^{2t\,P\cdot X/R}  \label{SchwingerRep}
\ ,
\ee
where
\be
\mathcal{C}_\Delta = \frac{\Gamma(\Delta)}{2\pi^h \Gamma\left(\Delta-h+1\right)}\ , \ \ \ \ \ \ \ \ \ \ \ \
h=\frac{d}{2}\ . \label{btobprop}
\ee
This normalization was obtained from taking the limit of the bulk to bulk propagator
\cite{KlebanovWitten}.
This gives rise to the following normalization of the two-point function
\be
\langle \mathcal{O}_\Delta(P_1)  \mathcal{O}_\Delta(P_2) \rangle = \frac{\mathcal{C}_\Delta}
{(-2P_1 \cdot P_2)^\Delta}\ . \label{AdSnormalization}
\ee

One can also describe tensor fields in AdS using this language.
A tensor field in AdS can be represented by a transverse tensor field in $\mathbb{M}^{d+2}$,
\be
X^{A_i}T_{A_1\dots A_l}(X)=0\ .
\ee
Covariant derivatives in AdS can be easily obtained from simple partial derivatives
in the flat embedding space.
The rule is to take partial derivatives of transverse tensors and then project into the tangent space of AdS using the projector
\be
U^A_B=\delta_A^B + \frac{X_AX^B}{R^2}\ .  \label{projector}
\ee
For example
\ba
\nabla_{A_3}\nabla_{A_2}T_{A_1}(X) = U^{B_3}_{A_3} U^{B_2}_{A_2} U^{B_1}_{A_1}\,\partial_{B_3}
\left( U^{C_2}_{B_2}U^{C_1}_{B_1}\,\partial_{C_2}T_{C_1}(X) \right)\ . \label{covder}
\ea


\subsection{Contact interaction}
\label{sectioncontact}

\begin{figure}
\begin{center}
\includegraphics[
height=5cm]{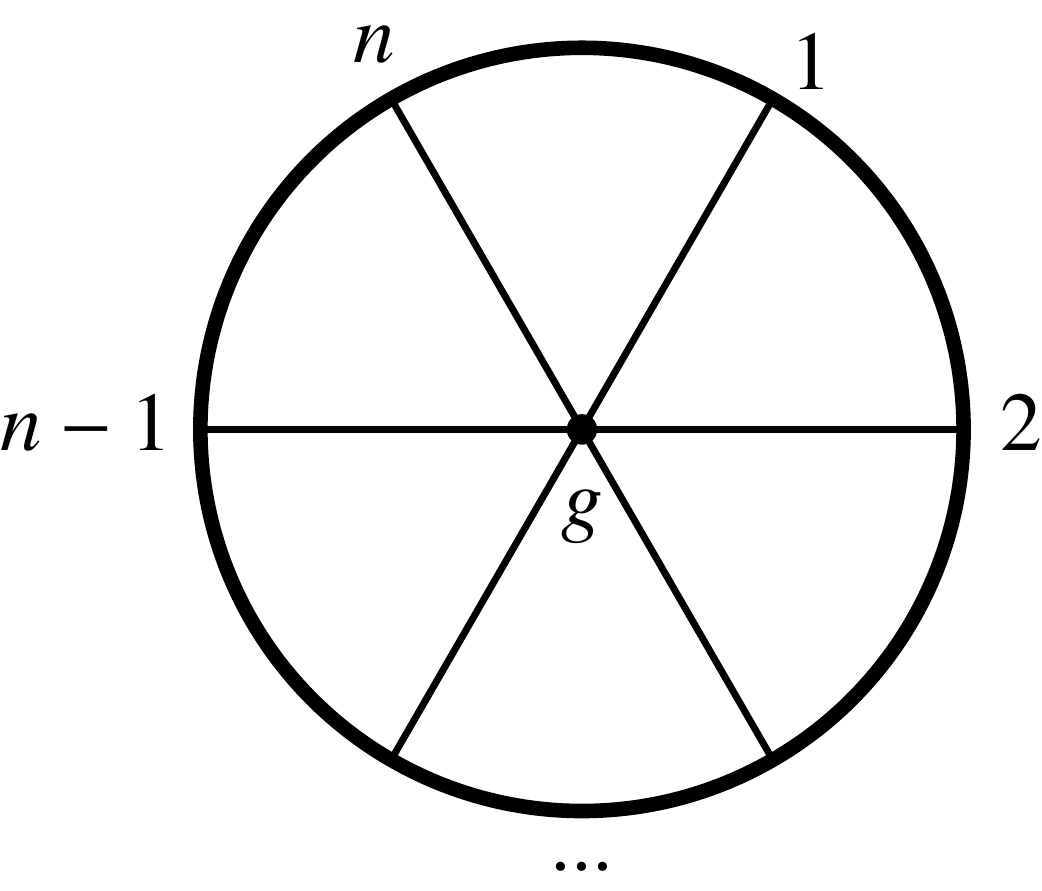}
\end{center}
\caption{ \small Witten diagram for a tree level $n$-point contact interaction in AdS. }
\label{contactfig}
\end{figure}

Let us start by considering the simple Witten diagram in figure \ref{contactfig},
\be
A(P_i)=g\,\int_{\rm AdS} dX \prod_{i=1}^n G_{B\partial}(X,P_i)\ ,
\ee
where $g$ is a coupling constant.
Using the representation (\ref{SchwingerRep}), we obtain the following expression for the $n$-point function
\be
A(P_i)=g R^{n(1-d)/2+d+1}\left(\prod_{i=1}^n \mathcal{C}_{\Delta_i}\right) \,D_{\Delta_1\dots\Delta_n}(P_i)\ , \label{contactWitten}
\ee
where we introduced the D-functions \footnote{We define the D-functions with the normalization of \cite{D'Hoker}.}
\be
D_{\Delta_1\dots\Delta_n}(P_i)=\prod_{i=1}^n \frac{1}{\Gamma(\Delta_i)} \int_0^\infty \frac{dt_1}{t_1}t_1^{\Delta_1} \dots
\int_0^\infty \frac{dt_n}{t_n}t_n^{\Delta_n} \int_{\rm AdS} d(X/R) e^{-2 Q\cdot X/R}
  \ . \label{Dfunction}
\ee
Here, $Q=\sum_{i=1}^n t_i P_i$ is a future directed vector in $\mathbb{M}^{d+2}$.
As explained in appendix \ref{appScalar}, parametrizing AdS with Poincare coordinates it is easy to show that
\be
\int_{\rm AdS} d(X/R) e^{-2 Q\cdot X/R}=\pi^{h} \int_0^\infty \frac{dz}{z} z^{-h} e^{-z+Q^2/z}\ ,
\ee
where we recall that $h=d/2$.
Rescaling $t_i \to t_i \sqrt{z}$ in (\ref{Dfunction}) the integral over $z$ factorizes and we obtain
\be
A(P_i)=2g R^{n(1-d)/2+d+1}\Ncal
\int_0^\infty \frac{dt_1}{t_1}t_1^{\Delta_1} \dots
\int_0^\infty \frac{dt_n}{t_n}t_n^{\Delta_n} e^{-\sum_{i<j} t_i t_j P_{ij}}\ ,
\ee
where we introduced the normalization constant $\Ncal$ given by
\be
2\Ncal=\pi^h\Gamma\left(\frac{\sum_{i=1}^n \Delta_i -d}{2} \right)\prod_{i=1}^n \frac{\mathcal{C}_{\Delta_i}}{\Gamma(\Delta_i)} \ .
\ee
In \cite{Symanzik} Symanzik showed that this integral can be written in the form (\ref{genMellin}) with
\be
M(\delta_{ij})=g R^{n(1-d)/2+d+1}\ . \label{norma}
\ee
We conclude that, as in flat space, tree level contact graphs in AdS give constant scattering amplitudes.
Moreover, we chose the normalization constant $\mathcal{N}$ so that
the contact interaction in AdS has the simplest possible Mellin amplitude.
We remark that the power of the AdS radius $R$ in (\ref{norma}) is the correct one to give a dimensionless Mellin amplitude.

We can also consider non-minimal contact diagrams, where the vertex includes covariant derivatives.
Take for example the interaction vertex $g_1 \nabla_A \phi_1 \nabla^A \phi_2 \,\phi_3 \dots \phi_n$,
where $\phi_i$ is the bulk scalar field dual to the operator $\Ocal_i$.
Using the rule (\ref{covder}) one can easily compute the associated $n$-point function
\be
A(P_i)=g_1 R^{n(1-d)/2+d-1}\left(\prod_{i=1}^n \mathcal{C}_{\Delta_i}\right)
  \Delta_1\Delta_2
\left[ D_{\Delta_1\dots\Delta_n}(P_i) -2P_{12} \,D_{\Delta_1+1\, \Delta_2+1\, \Delta_3\dots\Delta_n}(P_i)\right]\ .
\ee
It is also easy to see that a generic interaction vertex with $2N$ covariant derivatives will
give rise to a $n$-point function that can be written as a linear combination of terms like
\be
 D_{ \Delta_1+\Lambda_1 \, \dots\,\Delta_n+\Lambda_n} (P_i)\,\prod_{i<j}^n P_{ij}^{\lambda_{ij}}
\ee
where $\Lambda_i$ and $\lambda_{ij}$ are non-negative integers obeying $\Lambda_i=\sum_{j\neq i} \lambda_{ij}$ and $\sum_{i<j} \lambda_{ij} \le N$.
The Mellin representation $M(\delta_{ij})$ for each of these terms is proportional to
\be
  \frac{\left(\frac{1}{2}\sum_i \Delta_i -h\right)_{\sum_{i<j} \lambda_{ij} }
}{\prod_{i}(\Delta_i)_{\Lambda_i}}
 \prod_{i<j}  (\delta_{ij})_{\lambda_{ij} }\ ,\label{term}
\ee
where we used the Pochhammer symbol $(a)_b=\Gamma(a+b)/\Gamma(a)$.
This function is a polynomial of degree $\sum_{i<j} \lambda_{ij}$ in the variables $\delta_{ij}$.
Therefore, we conclude that an AdS contact interaction with $2N$ covariant derivatives produces a polynomial Mellin amplitude
$M(\delta_{ij})$ of degree $N$.
As before, there is a striking similarity with flat space scattering amplitudes as functions of the Mandelstam invariants.

Let us try to make this similarity more precise.
Consider the same interaction vertex in flat spacetime.
The vertex contains $\alpha_{ij}$ pairs of contracted derivatives acting on the field
$\phi_i$ and the field $\phi_j$.
The total number of derivatives is then $2\sum_{i<j}^n \alpha_{ij}=2N$.
Assuming that all particles are massless, the flat space scattering amplitude is
\be
T(S_{ij})= g_N \prod_{i<j}^n \left(\frac{S_{ij}}{2}\right)^{\alpha_{ij}}\ ,
\ee
where $S_{ij}=-(K_i+K_j)^2=-2K_i\cdot K_j$ are the Mandelstam invariants, and $K_i$ is the momentum of particle $i$.
Let us now compare this result with the large $\delta_{ij}$ limit of the $M(\delta_{ij})$ associated with the same
interaction vertex in AdS.
From (\ref{term}) it is easy to see that the terms that dominate in the large $\delta_{ij}$ limit
are the ones with maximal $\sum_{i<j} \lambda_{ij}= N$.
Then, the rule (\ref{covder}) for the covariant derivative implies that this term is obtained simply by dropping
the projector (\ref{projector}). This gives
\be
A(P_i)\approx g_N R^{n(1-d)/2+d+1-2N}(-2)^N 
\left(\prod_{i=1}^n \mathcal{C}_{\Delta_i} (\Delta_i)_{\Lambda_i}\right)
D_{ \Delta_1+\Lambda_1 \, \dots\,\Delta_n+\Lambda_n} (P_i)\,\prod_{i<j}^n P_{ij}^{\alpha_{ij}}\ ,
\ee
where $\Lambda_i=\sum_{j\neq i} \alpha_{ij}$.
The large $s_{ij}$ limit of $M(s_{ij})$ is then given by
\be
M(s_{ij})\approx g_N R^{n(1-d)/2+d+1-2N} 
\frac{\Gamma\left(\frac{1}{2}\sum_i \Delta_i -h+N\right)}
{\Gamma\left(\frac{1}{2}\sum_i \Delta_i -h \right)}
\, \prod_{i<j}^n (s_{ij})^{\alpha_{ij}}\ .
\label{Mellincontactgen}
\ee
This suggests the following general relation
\be
M(s_{ij})\approx \frac{R^{n(1-d)/2+d+1}}{\Gamma \left(\frac{1}{2}\sum_i \Delta_i -h\right)}
\int\limits_0^\infty  d\beta  \beta^{\frac{1}{2}\sum_i \Delta_i -h-1}e^{-\beta} 
T\left(S_{ij}=\frac{2  \beta }{R^2} s_{ij}\right)\ ,
\ee
where the role of the integral is simply to produce the $N$-dependent
$\Gamma$-function in (\ref{Mellincontactgen}).
Notice that the powers of $R$ are consistent with dimensional analysis.
We have just shown that (\ref{mainformula}) is valid for all contact interactions
with arbitrary number of derivatives.
This is a very large class of interactions in the sense that 
other types of diagrams, like exchange diagrams, can be 
thought as infinite sums of these. 
This strongly suggest that  (\ref{mainformula}) is valid in general.
We shall find further evidence for this relation in the following sections.
We guessed (\ref{mainformula})
 using polynomial amplitudes but, in general, the scattering 
amplitude $T$ will have singularities and discontinuities.
These will give rise to singularities and discontinuities of the Mellin amplitude.
 
 Finally, we can invert (\ref{mainformula}) and obtain
\be
T(S_{ij})=  \Gamma \left(\frac{1}{2}\sum_i \Delta_i -h\right)
 \lim_{R \to \infty}
\int\limits_{ -i \infty}^{ i\infty} \frac{d\alpha}{2\pi i} \alpha^{h-\frac{1}{2}\sum_i \Delta_i  }e^{\alpha}
\frac{M\left(s_{ij}= \frac{ R^2}{2\alpha}S_{ij}\right)}
{R^{n(1-d)/2+d+1}}
\ ,
\label{invmainformula}
\ee
where the integration contour in the $\alpha$-plane passes to the right  of all poles of the integrand.

\subsection{Scalar and graviton exchange}
\label{sectionScaExc}

\begin{figure}
\begin{center}
\includegraphics[
height=5cm]{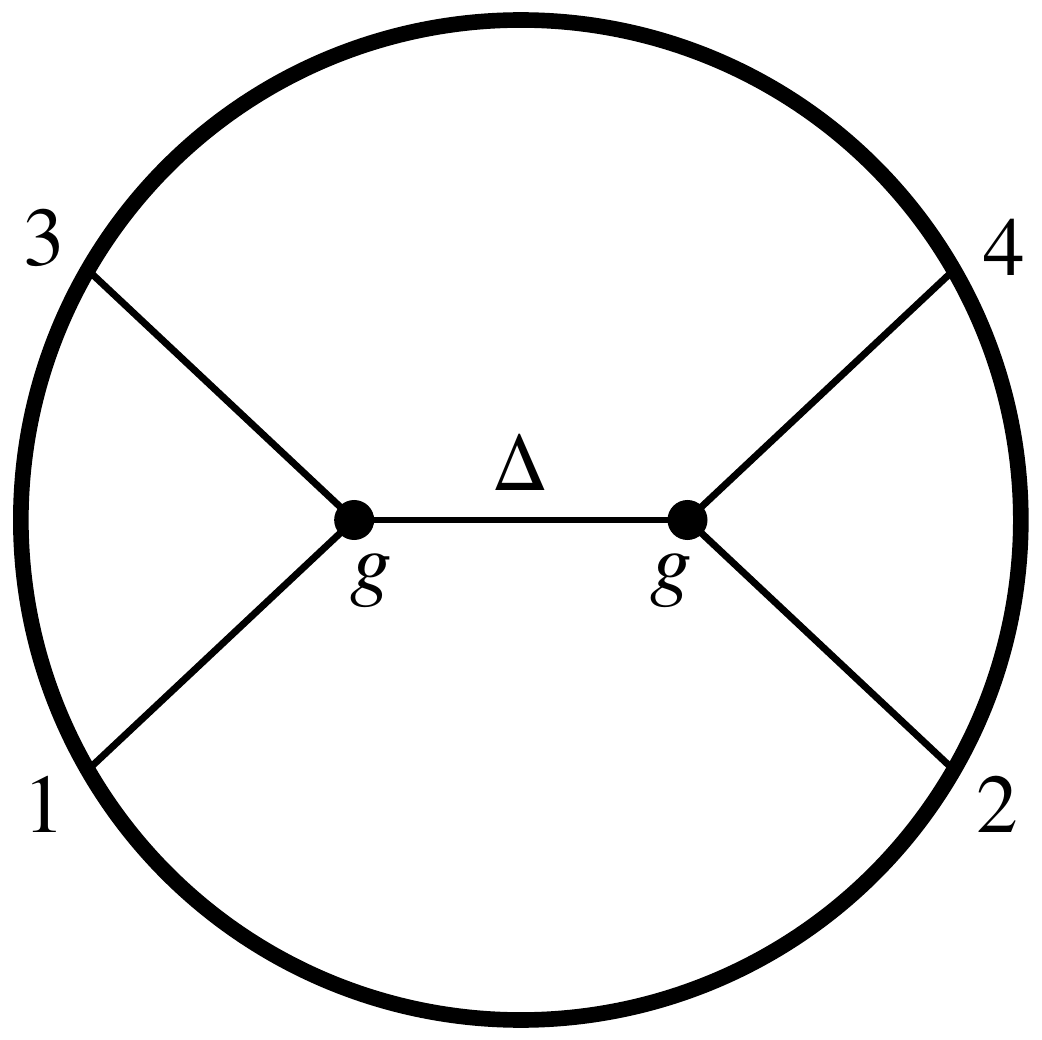}
\end{center}
\caption{\small Witten diagram for a tree level scalar exchange in AdS. }
\label{4ptexchangefig}
\end{figure}

Consider the 4pt function associated with the scalar exchange diagram of figure \ref{4ptexchangefig}.
In the special case where the dimension $\Delta$ of the operator dual to the exchanged scalar satisfies $m=(\Delta_1+\Delta_3-\Delta)/2 \in \mathbb{N}$,
 the authors of \cite{Howtozintegrals} reduced this diagram to the following sum of D-functions
\be
A(P_i)=g^2 R^{5-d} \prod_{i=1}^4 \mathcal{C}_{\Delta_i}  \sum_{l=1}^{m} a_l ( P_{13})^{-l}
D_{\Delta_1-l\,\Delta_2  \,\Delta_3-l \,\Delta_4 } (P_i)\ ,
\ee
with
\be
a_l= \frac{(\Delta_1)_{-l} (\Delta_3)_{-l}}{4
   \left(\frac{\Delta_1+\Delta_3-\Delta}{2}  \right)_{1-l} \left(\frac{\Delta_1+\Delta_3-d+\Delta}{2}
  \right)_{1-l}}\ .
\ee
This gives the Mellin amplitude
\begin{align}
M(\delta_{ij})&=g^2 R^{5-d}\sum_{l=1}^{m} a_l \frac{ \left(\frac{\sum_i \Delta_i -d}{2}\right)_{-l} (\delta_{13})_{-l} }{(\Delta_1)_{-l} (\Delta_3)_{-l}}\nonumber \\
  &=g^2 R^{5-d}\frac{\,_3F_2\left(1,1+\frac{s_{13}-\Delta }{2},1+\frac{s_{13}+\Delta-d }{2};
2+ \frac{s_{13}- \Delta_1-\Delta_3}{2},2+ \frac{s_{13} -\Delta_2-\Delta_4}{2}  ;1 \right)}{
\left(2+  s_{13}- \Delta_1-\Delta_3 \right)\left(2+  s_{13}- \Delta_2-\Delta_4 \right)} \ .
\label{genscalar}
\end{align}
This result was derived assuming that $(\Delta_1+\Delta_3-\Delta)/2$ was a non-negative integer but  the final expression is valid for general $\Delta$ as we show in appendix \ref{appScalar} by directly computing the diagram.
Notice that the Mellin amplitude only depends on $s_{13}$ as expected
for a scalar exchange.
\footnote{
For massless fields  in AdS$_5$ ($\Delta_i=\Delta=d=4$)  we obtain the  simple result
\be
M(s_{ij})=g^2 R^{5-d} \frac{11-2  s_{13}}{20 (s_{13}-6) ( s_{13}-4)}\ . \nonumber
\ee
Notice that $ M(s_{ij})\approx  - g^2 R^{5-d} /(10 s_{13} )$
for large $s_{13}$ in perfect agreement with the general formula (\ref{mainformula}) if we use $T(S_{ij})=-g^2/S_{13}$ for a massless
scalar exchange in flat space.}

We now wish to study the analytic structure of  (\ref{genscalar}).
In order to do that, it is convenient to use the following Mellin-Barnes representation,
which we derive in appendix \ref{appScalar},
\be
M(s_{ij})=
\frac{g^2 R^{5-d}}{
\Gamma\left(\frac{\sum_i\Delta_i }{2}-h \right)
\Gamma\left( \frac{\Delta_1+\Delta_3-s_{13}}{2} \right)
\Gamma\left( \frac{\Delta_2+\Delta_4-s_{13}}{2} \right)
}
\int\limits_{-i\infty}^{i\infty} \frac{dc}{2\pi i }
 \frac{l(c)l(-c)}{(\Delta-h)^2-c^2 }\ ,\label{MBrep}
\ee
where
\be
l(c)=\frac{\Gamma\left(\frac{ h+c-s_{13}}{2} \right)
\Gamma\left(   \frac{\Delta_1+\Delta_3+c-h}{2} \right)
 \Gamma\left(\frac{\Delta_2+\Delta_4+c-h}{2} \right)}{2 \Gamma(c)}\ . \label{lofc}
\ee
Poles in $s_{13}$ arise  from pinching of the integration contour in (\ref{MBrep})
between two colliding poles of the integrand.
In fact, we can write
\be
M(s_{ij})=-g^2 R^{5-d} \sum_{m=0}^\infty \frac{R_m}{s_{13}-\Delta-2m}\ , \label{sumM}
\ee
with
\be
R_m=
\frac{\Gamma \left(\frac{\Delta_1+\Delta_3+\Delta-d}{2}  \right)
   \Gamma \left(\frac{\Delta_2+\Delta_4+\Delta-d}{2}  \right)
}{2   \Gamma \left( \frac{\sum_i \Delta_i  -d  }{2}\right)   }
  \frac{ \left(1+\frac{\Delta-\Delta_1-\Delta_3}{2} \right)_m
     \left(1+\frac{\Delta-\Delta_2-\Delta_4}{2}  \right)_m}{m!  \Gamma\left(\Delta-\frac{d}{2}+1+m\right) }\ .
\ee
The poles in $s_{13}$ in (\ref{sumM}) appear exactly as predicted in (\ref{polesinMellin}).

We shall now consider the flat space limit $R \to \infty$ keeping the mass $\Delta(\Delta-d)/R^2$ of the exchanged particle
finite.
We want to know the value of the integral (\ref{MBrep}) for large $s_{13}$ and $\Delta^2$ of the same order.
One suggestive way of achieving this is to write $c=i K R$ and take the limit $R \to \infty$ with fixed $K$, $s_{13}/R^2$ and $\Delta^2/R^2$.
It is important that we consider $s_{13}$ large and away from the positive real axis where the Mellin amplitude has poles. A convenient choice is to consider negative $s_{13}$.
Using the Stirling expansion of the $\Gamma$-function we find
\be
 \frac{l(iKR) l(-iKR)}{
\Gamma\left( \frac{\Delta_1+\Delta_3-s_{13}}{2} \right)
\Gamma\left( \frac{\Delta_2+\Delta_4-s_{13}}{2} \right)} \approx \frac{2\pi}{|K | R }  \left(-\frac{K^2R^2}{2s_{13}} \right)^{\frac{\sum \Delta_i}{2}-h} e^{\frac{K^2R^2}{2s_{13}}}\ ,
\ee
for large $R$, with fixed $K$, $\Delta^2/R^2$ and $s_{13}/R^2<0$.
Thus
\be
M(s_{ij}) \approx
\frac{ 2R^{3-d}}{
\Gamma\left(\frac{1}{2}\sum_i\Delta_i  -h \right) }
\int\limits_{0}^{\infty} \frac{dK}{ K }
  \left(-\frac{K^2R^2}{2s_{13}} \right)^{\frac{1}{2}\sum_i\Delta_i -h} e^{\frac{K^2R^2}{2s_{13}}}   \frac{g^2}{\Delta^2/R^2+K^2}\ ,
\ee
where we have used the invariance of the integrand under $K \to -K$.
The last expression becomes 
 \be
M(s_{ij})
\approx \frac{R^{3-d}}{    \Gamma \left(\frac{1}{2}\sum_i\Delta_i -h\right)}  \int\limits_0^\infty   d\beta \,
\beta^{\frac{1}{2}\sum_i\Delta_i -h-1}\,
e^{-\beta}  \frac{g^2 }{-2 s_{13}\beta/R^2+\Delta^2/R^2   }\ ,
 \label{FSLexchange}
\ee
 after the change of integration variable $K^2=-2\beta  s_{13}/R^2$.
 This is in perfect agreement with the general formula (\ref{mainformula}) for the case
of a massive scalar exchange with mass $\Delta^2/R^2$.

The integral (\ref{FSLexchange}) can be written in terms of the standard
exponential integral function,
\be
M(s_{ij}) \approx \frac{-g^2R^{5-d}}{2s_{13}} e^{-\frac{\Delta^2}{2s_{13}}}
\, {\rm Ei}_{\frac{1}{2}\sum_i \Delta_i  -h} \left( -\frac{\Delta^2}{2s_{13}} \right)\ .
\label{FSLexchangeexp}
\ee
The only singularities of the  Mellin amplitude (\ref{MBrep}) are a series of single poles on the positive real axis as shown in (\ref{sumM}). However, in the limit of large $s_{13}$, these poles condense and generate 
a branch cut along the positive real axis of $s_{13}$ in (\ref{FSLexchangeexp}).
Notice that the origin of this discontinuity in expression (\ref{FSLexchange}) was   the pole of the scattering amplitude $T(S_{13})$ at $S_{13}=\Delta^2/R^2$.

\begin{figure}
\begin{center}
\includegraphics[
height=5cm]{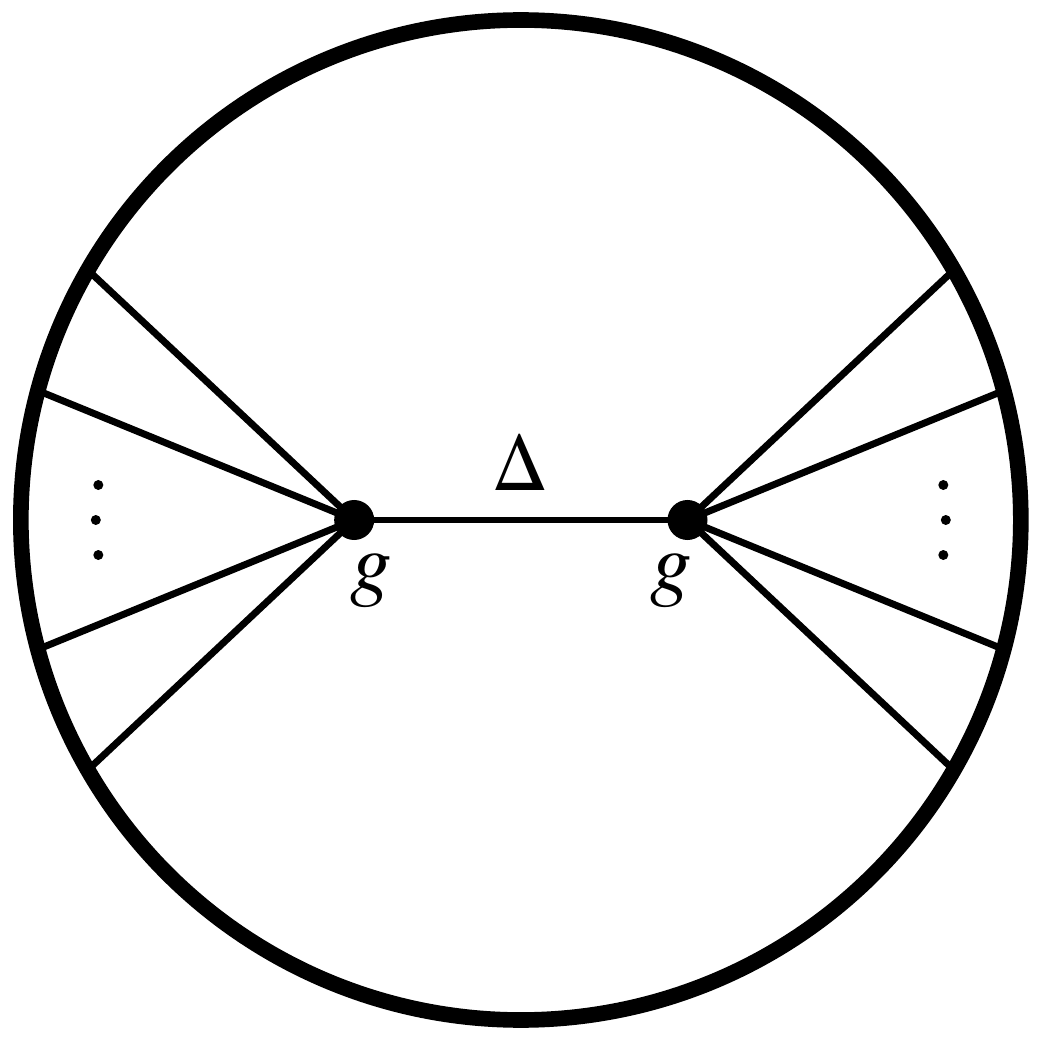}
\end{center}
\caption{ \small A tree level scalar exchange in AdS contributing to a $n$-point correlation function. }
\label{nptexchangefig}
\end{figure}

Another instructive example is the Witten diagram in figure \ref{nptexchangefig}.
The computation of its associated Mellin amplitude is almost identical to the previous example if one follows the method explained in appendix \ref{appScalar}.
The result is given by
\be
M(s_{ij})=
\frac{g^2 R^{5-d}}{
\Gamma\Big(\frac{1}{2}\sum\limits_i\Delta_i -h \Big)
\Gamma \Big( \sum\limits_{i,j\in \mathcal{L}}  \delta_{ij} \Big)
\Gamma \Big(  \sum\limits_{i,j\in \mathcal{R}}  \delta_{ij}  \Big)
}
\int\limits_{-i\infty}^{i\infty} \frac{dc}{2\pi i }
\frac{ l(c)l(-c) }{(\Delta-h)^2-c^2}  \ ,
\ee
where $\mathcal{L}$ ($\mathcal{R}$) is the group of points that connect to the left (right) interaction vertex in figure
\ref{nptexchangefig}, and
\be
l(c)=\frac{\Gamma\Big(\frac{ h+c}{2}-\frac{1}{2}\sum\limits_{i\in \mathcal{L}}\sum\limits_{j\in \mathcal{R}} \delta_{ij}  \Big)
\Gamma\Big(  \frac{c-h}{2}+\frac{1}{2}\sum\limits_{i\in \mathcal{L}}\Delta_i  \Big)
 \Gamma\Big(\frac{c-h}{2}+\frac{1}{2}\sum\limits_{i\in \mathcal{R}}\Delta_i \Big)}{2 \Gamma(c)}\ .
\ee
This shows that the only poles of the Mellin amplitude are at
\be
\sum_{i\in \mathcal{L}}\sum_{j\in \mathcal{R}} \delta_{ij}=\Delta+2m \ , \ \ \ \ \ \ \ \ \ m=0,1,2,\dots\ .
\ee
Let us introduce "momentum" $k_i$ associated with operator $\Ocal_i$, such that $-k_i^2=\Delta_i$ and $\sum_i k_i=0$.
Then, if we write $\delta_{ij}=k_i\cdot k_j$ as in (\ref{deltatos}), the pole condition reads
\be
\sum_{i\in \mathcal{L}}\sum_{j\in \mathcal{R}} \delta_{ij}=
\sum_{i\in \mathcal{L}}\sum_{j\in \mathcal{R}} k_i\cdot k_j=
-\Big(\sum_{i\in \mathcal{L}} k_i\Big)^2=\Delta+2m \ , \ \ \ \ \ \ \ \ \ m=0,1,2,\dots
\ .
\ee
This has the suggestive interpretation of the total exchanged "momentum going
on-shell".

Finally, let us now return to the 
 graviton exchange process discussed in the introduction.
With our conventions, the Mellin amplitude associated with graviton exchange 
between minimally coupled massless scalars in AdS$_5$ ($\Delta_i=d=4$), is given by 
\be
M(s_{ij}) = -\frac{32\pi G_5 R^{-3}}{5} \left( 
\frac{6 \gamma_{13}^2+2}{s_{13}-2}+
\frac{8 \gamma_{13}^2}{s_{13}-4}+
\frac{\gamma_{13}^2-1}{s_{13}-6}
-\frac{15}{4} s_{13}+\frac{55}{2}
   \right)\ , \label{gravitonMellin}
\ee
where $G_5$ is the Newton's constant in AdS$_5$
and $\gamma_{13}=(s_{12}-s_{14})/2$.
The large $s_{ij}$ limit gives
\be
M(s_{ij})\approx  96\pi G_5 R^{-3} \frac{s_{12} s_{14}}{s_{13}}\ , 
\ee
in agreement with the result of formula (\ref{mainformula}) using
the scattering amplitude
\be
T(S_{ij})=8\pi G_5 \frac{S_{12}S_{14}}{S_{13}}\ ,
\ee
for graviton exchange between minimally coupled massless scalars in 
flat space \cite{FSgraviton1, FSgraviton2}.

\subsection{One-loop Witten diagram}
\label{sectiononeloop}

\begin{figure}
\begin{center}
\includegraphics[
height=5cm]{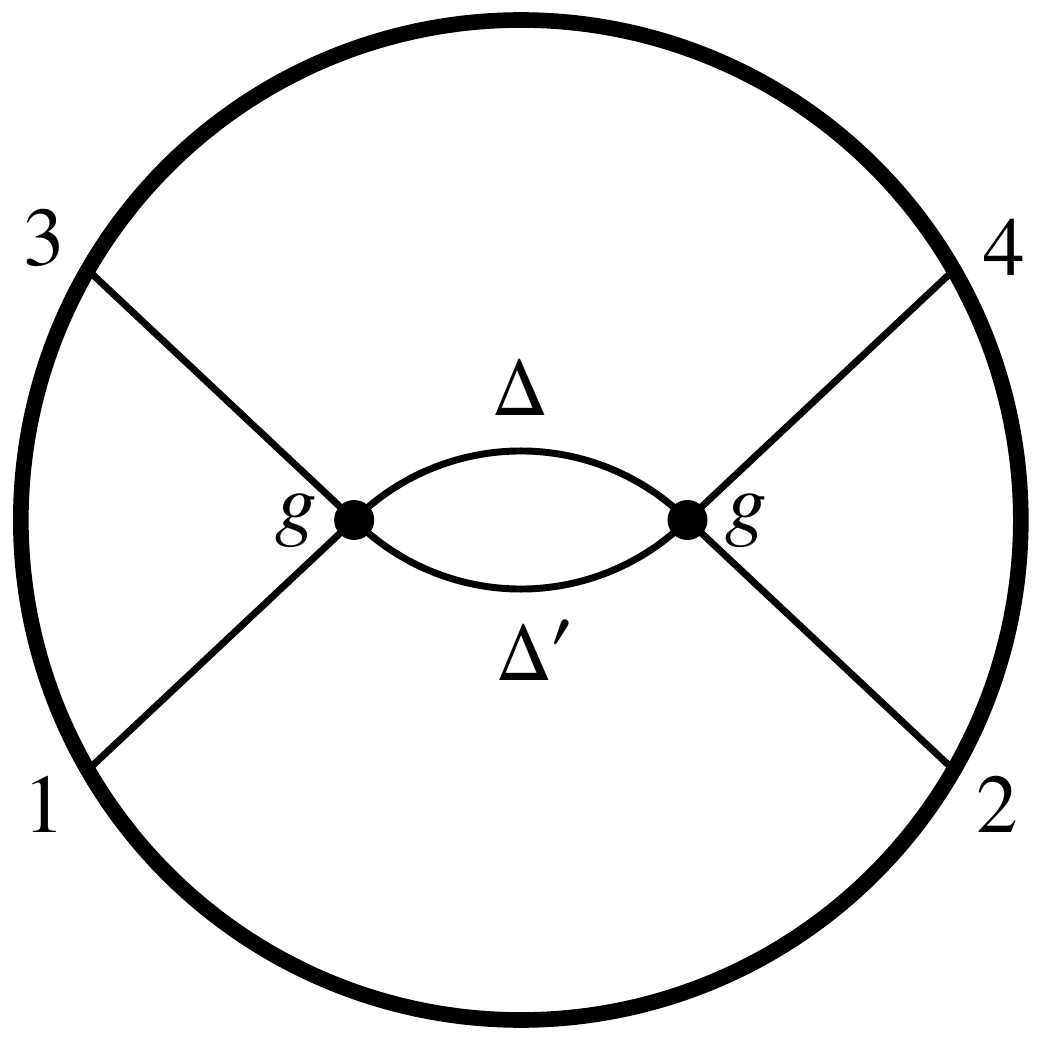}
\end{center}
\caption{ \small One-loop Witten diagram contributing to the $4$-point correlation function. }
\label{1loopfig}
\end{figure}

It is important to test our main formula (\ref{mainformula}) beyond tree level diagrams.
To this end, we shall study the 1-loop diagram  of figure \ref{1loopfig}.
The associated Mellin amplitude is computed in appendix \ref{appLoop}. The result reads
\be
M(s_{ij})=
\frac{g^2 R^{6-2d}}{
\Gamma\left(\frac{\sum_i\Delta_i }{2}-h \right)
\Gamma\left( \frac{\Delta_1+\Delta_3-s_{13}}{2} \right)
\Gamma\left( \frac{\Delta_2+\Delta_4-s_{13}}{2} \right)
}
\int\limits_{-i\infty}^{i\infty} \frac{dc}{2\pi i }
  l(c)l(-c) q(c ) \ ,\label{MB1loop}
\ee
where $l(c)$ is given by the same expression (\ref{lofc}) as in the tree level exchange and
\be
q(c )=\frac{\Gamma(c)\Gamma(-c)}{8\pi^h\Gamma(h)\Gamma(h+c)\Gamma(h-c)}\int\limits_{-i\infty}^{i \infty} \frac{dc_1dc_2}{(2\pi i)^2}
\frac{\Theta(c,c_1,c_2)}{\left((\Delta-h)^2-c_1^2\right)\left((\Delta'-h)^2-c_2^2\right)}\ , \label{qofc}
\ee
with
\be
\Theta(c_1,c_2,c_3) =\frac{ \prod_{\{\sigma_i=\pm\}} \Gamma\left(\frac{h+ \sigma_1c_1 + \sigma_2c_2+ \sigma_3c_3}{2}\right) }{ \prod_{i=1}^3 \Gamma(c_i)\Gamma(-c_i)  }\ .
\label{Theta}
\ee
Here, $\prod_{\{\sigma_i=\pm\}}$ denotes the product over the $2^3=8$ possible values of $(\sigma_1,\sigma_2,\sigma_3)$.
This one-loop Mellin amplitude is rather long but the fact that it is possible 
to write it down in such a closed form is remarkable.
Our goal with this example is simply to understand the singularity structure and the flat space limit of one-loop Mellin amplitudes.

\begin{figure}
\begin{center}
\includegraphics[
height=6cm]{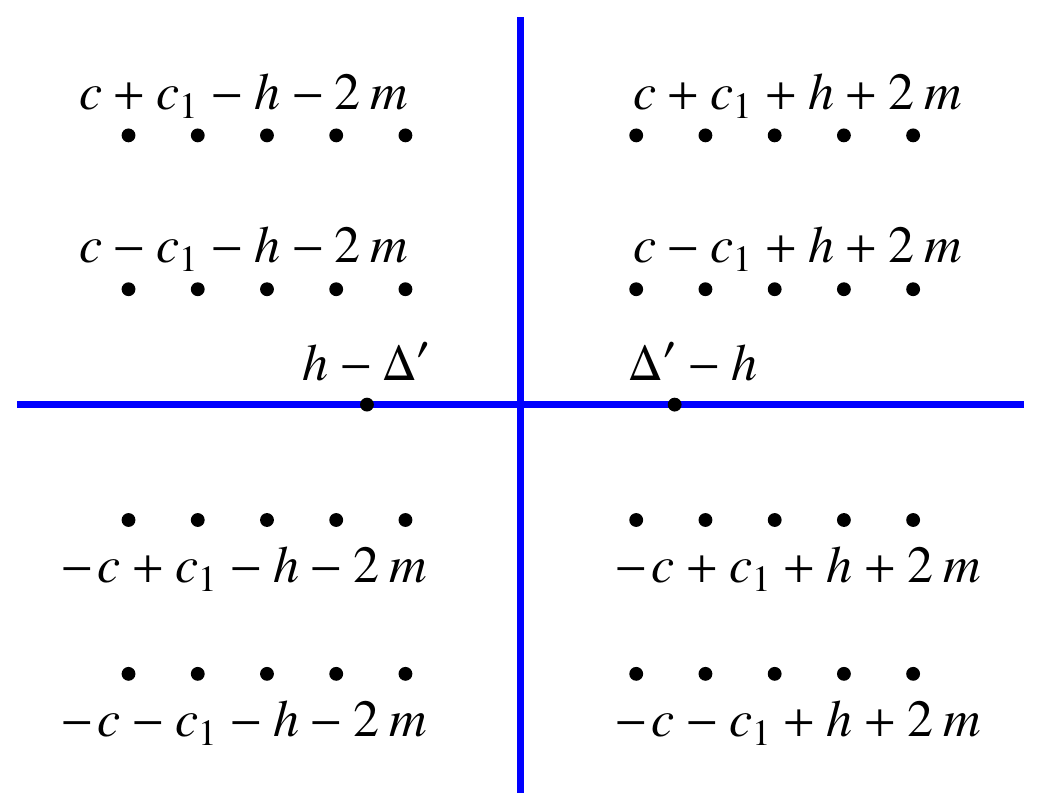}
\end{center}
\caption{ \small  Poles of the integrand in (\ref{qofc}) in the complex plane of $c_2$. There are 8 infinite sequences of poles labeled by non-negative integers $m$. The integration contour runs along the imaginary axis. }
\label{polesfig}
\end{figure}

The singularities of $M(s_{ij})$ are simple poles as for the tree level diagrams. This is a consequence of the 
discrete spectrum of a field theory in AdS. As before all poles are due to pinching of the integration contour between two colliding poles of the integrand. Let us start by finding the singularity structure of $q(c)$.
Firstly, we consider the integral over $c_2$ with fixed $c_1$.
The integrand has poles at
\be
\pm c_2= \Delta'-h\ ,\ \ \ \ \ \ \ \ \ \ \ \pm c_2= h \pm c \pm c_1 +2m\ ,\ \ \ \ \ \ \ \ \ m=0,1,2,\dots
\ee
where the several $\pm$ are uncorrelated, as shown in figure \ref{polesfig}. After performing the integral over $c_2$ we obtain a function of $c_1$ with poles at
\be
\pm c_1= \Delta'+2m \pm c\ ,\ \ \ \ \ \ \ \ \ \pm c_1=h+m\ ,\ \ \ \ \ \ \ \ \ m=0,1,2,\dots
\label{c1poles}
\ee
from pinching of the $c_2$-contour 
and
\be
\pm c_1= \Delta-h\ ,
\ee
from the explicit denominator in the integrand of (\ref{qofc}). 
One could also expect poles at $\pm c \pm c_1=h+m$ for $m=0,1,2,\dots$, from
pinching of the integration contour between reciprocal sequences of poles. However, these pole collisions happen at integer values of $c_2$ where the integrand has zeros from the factor $\Gamma(c_2)\Gamma(-c_2)$ in the denominator of (\ref{Theta}).
The integration over $c_2$ also generates poles at $\pm c=h+m$ for $m=0,1,2,\dots$, but these are canceled 
by the explicit factors of $\Gamma(h+c)$ and $\Gamma(h-c)$ in (\ref{qofc}).
Secondly, we perform the integral over $c_1$. 
Poles of $q(c)$ are generated by pinching of the $c_1$-integration contour between
the poles (\ref{c1poles}) and the poles $\pm c_1=\Delta-h$.
This gives poles at
\be
\pm c=\Delta+\Delta'-h +2m\ , \ \ \ \ \ \ \ \ \ m=0,1,2,\dots
\ee
Now that we know the poles of $q(c)$ the analysis is very similar to the tree level case (\ref{MBrep}).
The poles of the Mellin amplitude are at
\be
s_{13}=\Delta+\Delta' +2m\ , \ \ \ \ \ \ \ \ \ m=0,1,2,\dots
\ee
This corresponds precisely to the twist (i.e. conformal dimension minus spin) of the "double-trace" operators
\footnote{This is a schematic representation of the "double-trace" primary with spin $l$ and dimension $\Delta+\Delta'+2m+l$. In appendix \ref{DTapp}, we give the precise definition of this operator.}
\be
\mathcal{O}_{\Delta} \lrpar_{\mu_1} \dots \lrpar_{\mu_l} (\lrpar^{\nu}\lrpar_{\nu} )^m \mathcal{O}_{\Delta'}\ ,
\ee
dual to the two-particle states that are propagating between the two interaction vertices in figure \ref{1loopfig}.

The flat space limit of the Mellin amplitude (\ref{MB1loop}) can be obtained in a similar fashion to the tree level example 
considered in section \ref{sectionScaExc}.
More precisely, if we consider the change of variables $c=i K R$ and take the large $R$ 
limit with $s_{13}/R^2$  fixed and negative to avoid the poles, we obtain
\be
M(s_{ij}) \approx
\frac{ 2g^2R^{6-2d}}{
\Gamma\left(\frac{\sum_i\Delta_i }{2}-h \right) }
\int_0^{\infty} \frac{dK}{K }
  \left(-\frac{K^2R^2}{2s_{13}} \right)^{\frac{\sum \Delta_i}{2}-h} e^{\frac{K^2R^2}{2s_{13}}} q(i K R)\ .
\ee
In fact, matching with our general formula (\ref{mainformula})
for the flat space limit,
with the identification $K^2=-2\beta  s_{13}/R^2$, we conclude that 
\be
T(S_{13}=-K^2) = g^2 \lim_{R\to \infty} R^{3-d} q(i K R)\ ,
\ee
where $T(S_{13})$ is the scattering amplitude for the corresponding 1-loop diagram in flat space
and the limit is taken keeping the masses $\Delta^2/R^2$ and $\Delta'^{2}/R^2$ of the internal particles fixed.
 
To check this prediction, we change integration variables in (\ref{qofc}) as $c_1=iRK_1$ and $c_2=iRK_2$ and take the large $R$ limit of the integrand. Given the parity symmetry of the integrand it is enough to integrate over positive $K_1$ and $K_2$.
Using the Stirling approximation to the $\Gamma$-function we obtain the large $R$ behavior,
\be
\Theta(iRK_1,iRK_2,iRK_3) \approx e^{-\pi R \,\theta(K_1,K_2,K_3)} \frac{2\pi |K_1K_2K_3| }{ R^{1-4h}}   
\prod_{\{\sigma_i=\pm\}}\left|\sum_{i=1}^3\frac{ \sigma_i K_i}{2}\right|^{\frac{h-1}{2}}\ , \label{AsymTheta}
\ee
where
\be
\theta(K_1,K_2,K_3)=\frac{1}{4}\sum_{\{\sigma_i=\pm\}}|\sigma_1 K_1+\sigma_2 K_2+\sigma_3 K_3| -\sum_{i=1}^3 |K_i|\ ,
\ee
is never negative and it is zero if and only if it is possible to make a triangle
with sides $|K_1|$, $|K_2|$ and $|K_3|$. 
The exponential in (\ref{AsymTheta}) effectively cuts off the integration region over $K_1$ and $K_2$ and we obtain,
\be
g^2  \lim_{R\to \infty} R^{3-d} q(i K R) =  \frac{g^2}{ 4\pi^{h+1}\Gamma(h)}
 \int_{0}^{  \infty}  dK_1 dK_2
\frac{  K_1 K_2 \ 
\left(   Area(K_1,K_2,K)\right)^{2h-2}}{K^{2h-1}
  \left(\frac{\Delta^2}{R^2}+K_1^2\right)\left(\frac{\Delta'^2}{R^2}+K_2^2\right)} \ . \label{FSL1loopAdS}
\ee
where $Area(K_1,K_2,K_3)$ is the area of a triangle
with sides $|K_1|$, $|K_2|$ and $|K_3|$. It is zero if it is not possible to form
such a triangle and
\be
Area(K_1,K_2,K_3)=     
\left( \prod_{\{\sigma_i=\pm\}}\frac{\sigma_1 K_1+\sigma_2 K_2+\sigma_3 K_3}{2} \right) ^{\frac{1}{4}}
\ee
if it is possible.

This should be compared to the expected flat space result
\be
T(S_{13}=-K^2) = g^2 \int_{\mathbb{R}^{d+1}} \frac{dK_1 dK_2}{(2\pi)^{2(d+1)}} 
\frac{(2\pi)^{d+1} \delta^{d+1}(K_1+K_2+K)} {\left(\frac{\Delta^2}{R^2}+K_1^2\right)
\left(\frac{\Delta'^2}{R^2}+K_2^2\right)}\ , \label{1loopFS}
\ee
where here $K_i$ denote  $(d+1)$-dimensional vectors.
The usual way to proceed is to eliminate one $(d+1)$-dimensional 
integral using the delta function and keep only one integral over the independent loop momentum.
However, in order to
recover the result we got from the flat space limit of AdS, what we will do is to integrate first
over all possible directions of the vectors $K_1$ and $K_2$ keeping their norm fixed.
More precisely, writing 
\be
 \int_{\mathbb{R}^{d+1}} dK_1dK_2 \to \int_0^\infty dK_1dK_2 K_1^d K_2^d \int_{S^{d}} d\hat{K}_1
 d\hat{K}_2
\ee
we see that expression (\ref{1loopFS}) turns into (\ref{FSL1loopAdS}) if 
\be
 \int_{S^{d}} d\hat{K}_1d\hat{K}_2  \delta^{d+1}(K_1\hat{K}_1 +K_2\hat{K}_2+K) =
 \frac{2 \pi^h}{\Gamma(h)} \frac{\left( 2 Area(K_1,K_2,K) \right)^{d-2}}{(K_1K_2K)^{d-1}}\ . \label{angintegral}
\ee
We derive (\ref{angintegral}) in appendix \ref{angularint}. This concludes the proof that the flat space limit formula (\ref{mainformula}) is valid in this one-loop example.

Loop diagrams are often divergent and require   renormalization. We should distinguish between IR and UV divergences.
Since UV divergences are local, they are the same in AdS and in flat space. In our example, the loop integral in (\ref{1loopFS}) is UV divergent for $d+1>4$. In AdS, the UV divergence of the Mellin amplitude comes from the large $c_1$ and $c_2$  integration region in (\ref{qofc}).
Therefore, it is still present in (\ref{FSL1loopAdS}), which we have just shown
that precisely gives the flat space result. In particular, the formula for 
the flat space limit is valid for dimensionally regularized amplitudes.
Infrared divergences are more subtle. 
All Witten diagrams are IR finite because the AdS radius $R$ acts as an IR cutoff.
Then, if a diagram is IR divergent in flat space, the flat space limit $R \to \infty$ of the corresponding Witten diagram in AdS
 gives a particular IR regularization of the flat space diagram.
Translating this regularization scheme to a more standard one like dimensional regularization is not obvious.
On the other hand, the AdS IR regularization is physically sensible and 
can be useful in some circumstances \cite{CallanWilczekAdS}.

The example of this section strongly suggests that the flat space limit of AdS/CFT encoded in the simple formula (\ref{mainformula}) works at  loop level
and for massless particles.

\section{Flat space limit of AdS}
\label{FSlimit}
The flat space limit   of  scattering processes in AdS has been studied previously \cite{Polchinski,Susskind,GGP,TakuyaFSL,Katz}. In particular,  \cite{TakuyaFSL} proposed an explicit relation between the CFT four point function and the bulk $2 \to 2$ scattering amplitude of the dual fields. The goal of this section is to show that this relation follows from our main formula (\ref{mainformula}).

Let us start by briefly reviewing the proposal of \cite{TakuyaFSL} for the case of $2 \to 2$
elastic  scattering of scalar particles. We assume that the bulk theory 
has an intrinsic length scale $\ell_s$ that remains finite in the flat space
 limit $R \to \infty$.
Then, the $(d+1)$-dimensional flat space scattering amplitude can be written as
\be
T(S_{ij})=\ell_s^{d-3} \, \mathcal{T}\left(g,\sin^2\frac{\theta}{2} , \ell_s^2 S_{12}\right)\ ,
\ee
where $\mathcal{T}$ is dimensionless, $g$ denotes all dimensionless parameters of the theory and $\theta$ is the scattering angle given by 
$\sin^2\frac{\theta}{2}=- S_{13}/S_{12}$.
The relation between this scattering amplitude and the CFT four point function of the dual operators is encoded in a specific Lorentzian kinematical limit.
More precisely, we define  the reduced four point function $\mathcal{A}$ 
by dividing $A$ by the disconnected correlator,  
\be
A(P_i)=\frac{\mathcal{C}_{\Delta_1} \mathcal{C}_{\Delta_2} }{
(P_{13}+i\epsilon)^{\Delta_1}(P_{24}+i\epsilon)^{\Delta_2}}
\mathcal{A}\left(g, \frac{R}{\ell_s} ,\sigma, \rho^2 \right)  \ ,
\ee
where we have introduced the appropriate $i\epsilon$ prescription for Lorentzian correlation functions.
The reduced four point function $\mathcal{A}$ depends on the dimensionless parameters $g$ that characterize the theory, on the ratio of the AdS radius $R$ to the intrinsic length scale   $\ell_s$
 and on two independent conformal invariants, which we choose to be
\be
\sigma^2=\frac{P_{13}P_{24}}{P_{12}P_{34}}\ ,\ \ \ \ \ \ \ \ \ \ \ \ 
\sinh^2\rho = \frac{\det P_{ij}}{4P_{13}P_{24}P_{12}P_{34}}\ ,
\label{sigmarho}
\ee
where the determinant is taken over $i$ and $j$.
If $P_{12}>0$, $P_{34}>0$ (spacelike) and $P_3$ and $P_4$ are inside the future 
lighcones of both $P_1$ and $P_2$, then the   scaling limit 
\be
\mathcal{F}(g,\sigma,\xi)=
\frac{1}{\sigma^{\Delta_1+\Delta_2}}
\lim_{R/\ell_s \to \infty}  \left( \frac{R}{\ell_s}  \right)^{d-2\Delta_1-2\Delta_2}
\mathcal{A}\left(g, \frac{R}{\ell_s} ,\sigma, \rho^2=-\frac{\ell_s^2}{R^2}\frac{1-\sigma}{\sigma}\xi^2 \right)  \ , \label{FlimitA}
\ee
is well defined.
The main result of \cite{TakuyaFSL} was to show that the flat space scattering amplitude is directly related to $\mathcal{F}$ via
\be
\mathcal{F}(g,\sigma,\xi)=\frac{(\pi \xi)^\frac{3-d}{2} }{
\mathcal{Q}\sqrt{\sigma(1-\sigma)}}
\int_0^\infty \frac{d\nu}{\nu} \left(\frac{\nu}{2}\right)^{2\Delta_1+2\Delta_2-\frac{d+3}{2}}
K_{\frac{d-3}{2}}(\xi \nu) i\mathcal{T}\left(g,\sigma ,\nu^2+i\epsilon\right)\ ,
\label{FfromT}
\ee
where
\be
\mathcal{Q}=\Gamma(\Delta_1)
\Gamma(\Delta_2)\Gamma(\Delta_1-h+1)\Gamma(\Delta_2-h+1)\ .
\ee
In the remainder of this section we show 
  that (\ref{FfromT}) follows from our main formula (\ref{mainformula}).

\subsection{Derivation of (\ref{FfromT})}

We start by rewriting (\ref{mainformula}) for the present case 
\be
M(s_{ij}) \approx  \frac{(R/\ell_s)^{3-d} }{\Gamma(\Delta_1+\Delta_2-h)} 
 \int_0^\infty \frac{d\beta}{\beta}  \beta^{\Delta_1+\Delta_2-h}
e^{-\beta }\mathcal{T}\left(g, -\frac{s_{13}}{s_{12}} ,2 s_{12}\beta 
\frac{\ell_s^2}{R^2} \right) \label{MfromT} 
\ee
In order to derive (\ref{FfromT}) from (\ref{MfromT}) we need to show that the small
$\rho$ behavior of the four point function is controlled by the large $s_{ij}$ behavior
of its Mellin amplitude.
To see this, we start from the definition of the Mellin amplitude of the four-point function,
\begin{align}
A(P_i)=\frac{\mathcal{N}}{(2\pi i)^{2}} \int   d^2 \delta_{ij}\,
M(\delta_{ij})
\prod_{i<j} \Gamma(\delta_{ij}) (P_{ij}+i\epsilon)^{-\delta_{ij}}\ , 
\label{LorMellin}
\end{align}
adapted to the Lorentzian regime.
The integration contour runs along the imaginary axis of $\delta_{ij}$ with $\Re \delta_{ij} >0$.  The constraints (\ref{constraint}), in the present case,
can be solved  by
\ba
\delta_{12}=    \delta_{34} =\epsilon-i s / 2  \ ,\ \ \ \ \ \ \ \ \ \ \ \ 
\delta_{13}=  \Delta_1-2\epsilon-it/2\ , \\
   \delta_{14}=    \delta_{23} =\epsilon-i u / 2  \ ,\ \ \ \ \ \ \ \ \ \ \ \ 
 \delta_{24}=\Delta_2-2\epsilon-it/2\ ,
\ea
where $s+t+u=0$ and $\epsilon>0$ is an infinitesimal parameter
important to give the correct integration contour. 
The integration measure is then given by
\be
\frac{1}{(2\pi i)^{2}} \int   d^2 \delta_{ij}\, (\dots) \to \frac{1}{(4\pi)^2}  \int dsdtdu\, \delta(s+t+u)\, (\dots)
\ee

\begin{figure}
\begin{center}
\includegraphics[
height=7cm]{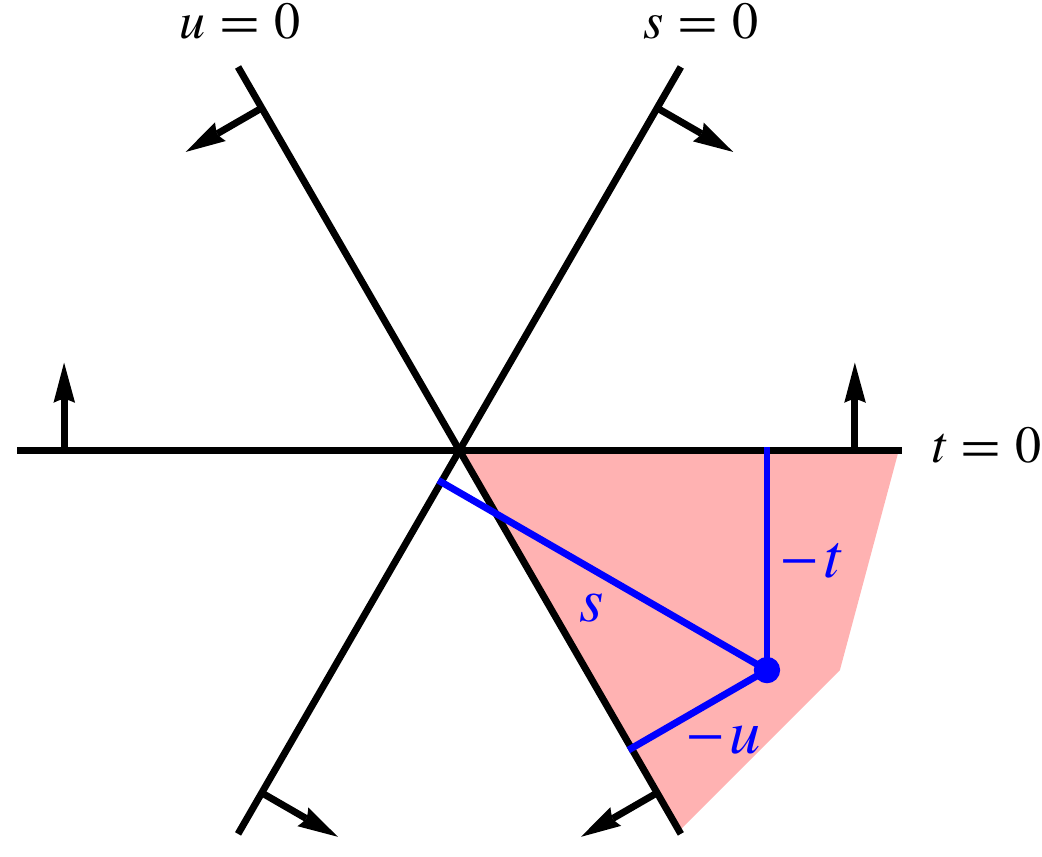}
\end{center}
\caption{ \small Parametrization of the two-dimensional integration surface. Any point on the plane has $s+t+u=0$. The region where the integrand of the Mellin representation of the Lorentzian four-point function does not decay exponentially is shown in light red. }
\label{STUplane}
\end{figure}

Using the asymptotic behavior of the $\Gamma$-function we find
\begin{align}
\prod_{i<j}\left| \Gamma(\delta_{ij})  \right|
\sim e^{-\frac{\pi}{2} (|s|+|t|+|u|)}
\end{align}
for large $s,\ t$ and $u$, up to power corrections.
In the Euclidean regime, this guarantees the convergence of the integral
in (\ref{LorMellin}) if $M(\delta_{ij})$ does not grow exponentially fast
at  infinity. Notice that, in the Euclidean regime  $P_{ij}>0$ and the factor
\begin{align}
 e^{ -\sum_{i<j} \delta_{ij}   \log(P_{ij}+i\epsilon) }
\end{align}
only gives an oscillating phase at large $s,\ t$ and $u$. 
Thus, the integrand decays exponentially in all directions of the
integration plane shown in figure \ref{STUplane}, due to the $\Gamma$-functions.
However, in the Lorentzian regime appropriate to study the flat space limit, we have $P_{13},P_{14},P_{23},P_{24}$ all negative.
This gives
\be
\left|e^{ -\sum_{i<j} \delta_{ij}   \log(P_{ij}+i\epsilon) } \right|  \sim \prod_{i<j} e^{\pi  \Theta(-P_{ij}) \Im \delta_{ij}}
\sim e^{\pi  s }
\ee
for the exponential decay at large $s,t$ and $u$.
In this case, the integrand does not decay exponentially in the "physical" region of positive $s$ and negative $t$ and $u$, as shown in figure \ref{STUplane}.
In fact, the singularity associated with the flat space limit can be determined from the region $s,-t,-u \gg 1$.
The integral over this region can be written as
\be
   \int dsdtdu\, \delta(s+t+u)\, (\dots)\to
    \int_0^\infty ds\, s \int_0^1d\eta \, (\dots)
\ee
with $t=-s \eta$ and $u=-s(1-\eta)$. We can also approximate the integrand for large $s$,
\ba
\prod_{i<j} \Gamma(\delta_{ij}) (P_{ij}+i\epsilon)^{-\delta_{ij}}
&\approx& -i \frac{(4\pi)^3 }{ stu} \exp \sum_{i<j} \delta_{ij} \Big[ \log (\delta_{ij})-1  - \log(P_{ij}+i\epsilon)   \Big]\\
&\approx& -i 
\frac{\left(\frac{4\pi }{ s}\right)^3 
\left( \frac{i\eta s}{2}\right)^{\Delta_1+\Delta_2} 
e^{-i\frac{s}{2}\left[\eta \log \frac{z\bar{z} }{ \eta^2}  +(1-\eta)\log\frac{(1-z)(1-\bar{z})}{(1-\eta)^2}\right]}}
{\eta(1-\eta)\,(P_{13}+i\epsilon)^{\Delta_1}(P_{24}+i\epsilon)^{\Delta_2}}
\ea
where we have introduced the following parametrization of the
conformal invariant cross-ratios
\be
z\bar{z}=\frac{P_{13}P_{24}}{P_{12}P_{34}}\ ,\ \ \ \ \ \ \ \ \ \ \ \ 
(1-z)(1-\bar{z})=\frac{P_{14}P_{23}}{P_{12}P_{34}}\ .
\ee
The two independent variables $z$ and $\bar{z}$ (not complex conjugate)
are related to the invariants $\sigma$ and $\rho$ introduced in 
(\ref{sigmarho}) by
\be
z=\sigma e^\rho\ ,\ \ \ \ \ \ \ \ \ \ \bar{z}=\sigma e^{-\rho}\ .
\ee
At large $s$, one can do the integral over $\eta$ by saddle point.
The stationary phase condition reads
\be
\frac{ \eta_s^2}{(1-\eta_s)^2}
= \frac{z\bar{z} }{(1-z)(1-\bar{z})}\ , 
\label{saddlecond}
\ee
and the expansion of the exponent around the saddle point is
\be
\eta \log \frac{z\bar{z} }{ \eta^2}  +(1-\eta)\log\frac{(1-z)(1-\bar{z})}{(1-\eta)^2}=
 \log \frac{z\bar{z} }{ \eta_s^2}- \frac{(\eta-\eta_s)^2}{\eta_s(1-\eta_s)}
 +O\left(( \eta-\eta_s)^3\right)\ .
\ee
This gives 
\be
\mathcal{A}\approx \frac{-i 4\pi  \mathcal{N}  }
{\mathcal{C}_{\Delta_1}\mathcal{C}_{\Delta_2}  \eta_s(1-\eta_s) }
\int_0^\infty \frac{ds}{s^2 } \left(\frac{i\eta_s s}{2}\right)^{\Delta_1+\Delta_2} 
\sqrt{\frac{2\pi\eta_s(1-\eta_s)}{-is}}e^{-i\frac{s}{2}  \log \frac{z\bar{z} }{ \eta_s^2} }
M(s_{ij})\ , \label{Aaftersaddle}
\ee
where $M(s_{ij})$ depends on $s$ and $\eta_s$ via $s_{12}=\Delta_1+\Delta_2+is\approx is$ and $s_{13}=it=-i\eta s\approx -i\eta_s s$, at the saddle point.
The reduced correlator $\mathcal{A}$ is enhanced  when the phase of the exponential factor in the integrand of (\ref{Aaftersaddle}) varies
slowly. This happens in the kinematical limit of small $\rho$.
To see this, one just needs to solve the saddle point condition (\ref{saddlecond}) at small $\rho$, 
\be
\eta_s=\sigma +\frac{\sigma^2 \rho^2}{2(1-\sigma)}+O(\rho^4) 
\ee
and replace into the exponential factor
\be
e^{-i\frac{s}{2}  \log \frac{z\bar{z} }{ \eta_s^2} } \approx e^{is\rho^2\frac{\sigma  }{2 (1-\sigma)} }\ .
\ee
This shows that the small $\rho$ behavior of the four point function
is controlled by the large $s_{ij}$ behavior of the Mellin amplitude.
At small $\rho$, we can then write 
\be
\mathcal{A}\approx \frac{-i 4\pi  \mathcal{N}  }
{\mathcal{C}_{\Delta_1}\mathcal{C}_{\Delta_2}  \sigma(1-\sigma) }
\int_0^\infty \frac{ds}{s^2 } \left(\frac{i\sigma s}{2}\right)^{\Delta_1+\Delta_2} 
\sqrt{\frac{2\pi\sigma(1-\sigma)}{-is}}e^{is\rho^2\frac{\sigma  }{2 (1-\sigma)}}
M(s_{ij})\ ,
\ee
where $M(s_{ij})$ depends on $s$ and $\sigma$ via 
$s_{12} \approx is$ and $s_{13} \approx -i\sigma s$.
 More precisely, we consider the limit $\rho \to 0$ with fixed $\xi$ given by
\be
\xi^2=-\rho^2 \frac{\sigma}{1-\sigma}\frac{R^2}{\ell_s^2}\ .
\ee
It is then natural to scale the integration variable $s\to s R^2/\ell_s^2$
to obtain
\be
\mathcal{A}\approx \frac{  4\pi^{\frac{3}{2}}  \mathcal{N}    }
{\mathcal{C}_{\Delta_1}\mathcal{C}_{\Delta_2}  \sqrt{ 1-\sigma} }
 \left(\frac{ i\sigma  }{2}\right)^{\Delta_1+\Delta_2-1/2} 
\int_0^\infty \frac{ds}{s  }  \left(\frac{R^2 }{\ell_s^2 } s \right)^{\Delta_1+\Delta_2-3/2}
e^{-is\frac{\xi^2}{2}} 
M(s_{ij})\ ,
\ee
where $M(s_{ij})$ is evaluated at $s_{12} \approx i s R^2/\ell_s^2$ and $s_{13}\approx-\sigma s_{12}$. 
We can now use (\ref{MfromT}) to replace the Mellin amplitude
by its approximate behavior at large $s_{ij}$,
\ba
\mathcal{A}&\approx& \frac{  4\pi^{\frac{3}{2}}  \mathcal{N}    }
{\mathcal{C}_{\Delta_1}\mathcal{C}_{\Delta_2}  
\Gamma(\Delta_1+\Delta_2-h)\sqrt{ 1-\sigma} }
  \left(\frac{R  }{\ell_s  }   \right)^{2\Delta_1+2\Delta_2-d }
 \left(\frac{ i\sigma  }{2}\right)^{\Delta_1+\Delta_2-1/2} 
 \\
 && 
\int_0^\infty \frac{ds}{s  }    s^{\Delta_1+\Delta_2-3/2}
e^{-is\frac{\xi^2}{2}}
 \int_0^\infty \frac{d\beta}{\beta}  \beta^{\Delta_1+\Delta_2-h}
e^{-\beta }\mathcal{T}\left(g, \sigma ,2i s \beta  \right) \ .
\ea
This shows that $\mathcal{A}$ has  the necessary scaling with $R/\ell_s$
to produce a well defined limit in (\ref{FlimitA}).
Moreover, after the rescaling $\beta \to \beta/(2s)$, we can perform the
integral over $s$ and obtain
\be
\mathcal{F} =  \frac{  (\pi\xi)^{\frac{3}{2}-h}     }
{  2 \mathcal{Q} \sqrt{ \sigma(1-\sigma)} } 
 \int_0^\infty \frac{d\beta}{\beta}  
 \left(\frac{i\beta}{4}\right)^{\Delta_1+\Delta_2-\frac{h}{2}-\frac{3}{4}}
 K_{h-\frac{3}{2}}\left(\xi\sqrt{i\beta}\right)\,
i\mathcal{T}\left(g, \sigma ,i   \beta  \right) 
\ ,
\ee
where $K$ is the modified Bessel function of the second kind.
Assuming that the scattering amplitude does not grow exponentially fast at large $S_{12}$ and that it is analytic for positive $\Re S_{12} $ and $\Im S_{12} $, the exponential decay of the Bessel function allows us to rotate the integration contour from $\beta  \in \mathbb{R}^+ $ to $i\beta  \in \mathbb{R}^+ $. Finally,  we can perform the change of variable $i\beta=\nu^2$ and  precisely recover the result (\ref{FfromT}).

\subsection{From SYM to strings in $\mathbb{M}^{10}$}

We can apply our result to the particular case of the 4pt-function of the Lagrangian density of $\Ncal=4$ super Yang-Mills.
The associated Mellin amplitude is the function
\be
M(g_{\rm YM}^2, \lambda , s_{ij})\ ,
\ee
where $\lambda=g_{\rm YM}^2N$ is the 't Hooft coupling.
Then, after including the contribution of the volume of 
the 5-sphere, our formula (\ref{invmainformula})  gives the full scattering amplitude for dilaton particles in type IIB superstring theory in $\mathbb{M}^{10}$,
\ba
T(g_s,\ell_s,S_{ij})&=&\Gamma(6)
 \lim_{R \to \infty} R\, {\rm volume}(S^5)
 \int_{ -i\infty}^{ i\infty} \frac{d\alpha}{2\pi i} \frac{e^\alpha}{\alpha^6}
 M\left( 4\pi g_s, \frac{R^4}{\ell_s^4} , \frac{S_{ij}R^2}{2\alpha} \right)
\\
&=& 120\pi^3 \ell_s^6 \lim_{\lambda \to \infty} \lambda^{3/2}
\int_{ -i\infty}^{ i\infty} \frac{d\alpha}{2\pi i} \frac{e^\alpha}{\alpha^6}
    M\left( 4\pi g_s, \lambda ,\sqrt{\lambda}  \frac{S_{ij}\ell_s^2 }{2\alpha} \right)\ .
 \ea
where we have used the standard relations
\be
4\pi g_s=g_{\rm YM}^2\ ,\ \ \ \ \ \ \ \ \ \ \ \ \ 
\left(\frac{R}{\ell_s}\right)^4=\lambda\ ,
\ee
between the string coupling $g_s$, the Yang-Mills coupling $g_{\rm YM}$, the   't Hooft coupling $\lambda$, the string length $\ell_s$,and the AdS radius $R$.
 
 In particular, if we focus on the planar four point function,
 we can use the type IIB superstring tree-level dilaton scattering amplitude to
 obtain the following constraint on the Mellin amplitude,
 \ba
&& \lim_{\lambda \to \infty} \lambda^{-1/2}
\int_{ -i\infty}^{ i\infty} \frac{d\alpha}{2\pi i} \frac{e^\alpha}{\alpha^6}
    M_{\rm planar}\left(s_{ij}=\sqrt{\lambda}  \frac{S_{ij}  }{2\alpha} \right)
  \label{FSLplanar}
    \\&=& \frac{1}{N^2} \frac{\pi^2}{30} 
    \left( \frac{S_{13}S_{14}}{S_{12}}+\frac{S_{12}S_{14}}{S_{13}}+
    \frac{S_{12}S_{13}}{S_{14}}\right) \mathcal{B}
    \left( \frac{S_{12}}{4}, \frac{S_{13}}{4}, \frac{S_{14}}{4}\right)\ ,
    \nonumber
 \ea
 where $S_{12}+S_{13}+S_{14}=0$  and
 \be
 \mathcal{B}(a_1,a_2,a_3)=\prod_{i=1}^3 \frac{\Gamma(1-a_i)}{\Gamma(1+a_i)}\ .
 \ee
 The leading term in the small $S_{ij}$ expansion of (\ref{FSLplanar})
 is a prediction for the Mellin amplitude in the supergravity approximation,
 \be
\int_{ -i\infty}^{ i\infty} \frac{d\alpha}{2\pi i} \frac{e^\alpha}{\alpha^6}
   \lim_{\lambda \to \infty}  
   \frac{  M_{\rm SUGRA}\left(s_{ij}=\sqrt{\lambda}  \frac{S_{ij}  }{2\alpha} \right)}{\sqrt{\lambda}}
 = \frac{1}{N^2} \frac{\pi^2}{30} 
    \left( \frac{S_{13}S_{14}}{S_{12}}+\frac{S_{12}S_{14}}{S_{13}}+
    \frac{S_{12}S_{13}}{S_{14}}\right) \ . \label{FSLsugra}
 \ee
 It was shown in \cite{D'Hoker} that the four point function of the Lagrangian density in the supergravity approximation is given 
 by the sum, over the 3 channels, of the graviton exchange process discussed
 at the end of section \ref{sectionScaExc}. 
 Therefore, the Mellin amplitude is a sum of 3 terms like (\ref{gravitonMellin}) corresponding to the 3 possible channels.
 Inserting this in (\ref{FSLsugra}) we obtain perfect agreement 
 using the relation $2 G_5 R^{-3} = \pi/ N^2$.

\section{Conclusion}
\label{conclusion}

Conformal correlation functions are rather complicated objects.
However, they are highly constrained by locality and the existence of the OPE. These constraints translate into crossing symmetry and meromorphy of the Mellin amplitudes. 
Moreover, in the case of conformal gauge theories in the planar limit,\footnote{More generally, CFT correlation functions dual to tree level processes in AdS gravitational theories.}
all poles of the Mellin amplitudes are associated with  single-trace operators. 
These properties make the Mellin amplitudes the ideal tools to attempt
the conformal bootstrap  program in higher dimensions. 
In particular, in $\mathcal{N}=4$ SYM the position of all poles is known since it is given by the spectrum
of local single-trace operators. It is tempting to imagine that the knowledge
of all singularities of the Mellin amplitude plus the constraints of crossing symmetry and factorization of the residues, completely fixes it.
A less ambitious approach is to try to construct four point functions from the knowledge of two and three point functions
of single-trace operators. Notice that, in general, this is only possible if all two and three point functions of primary operators are known, 
including multi-trace operators. However,
in the planar limit, all singularities (and their residues) of the Mellin amplitude  are fixed by the two (and three) point function of single-trace 
operators.

In the Mellin amplitudes the meaning of the CFT constraints is much more transparent. As an illustrative example, consider the problem studied in 
\cite{JP,IdseJamie}.
The main result of \cite{JP,IdseJamie} was to show that all consistent conformal four point functions of a single-trace operator $\mathcal{O}$ that does not contain any single-trace operator in the  $\mathcal{O}\mathcal{O}$ OPE,
are given by quartic contact graphs in AdS.
This result required a rather complicated analysis of the conformal partial wave decomposition.
On the other hand, absence of single-trace operators in the OPE translates into analyticity of   the Mellin amplitude. Moreover, in section \ref{sectioncontact} we
showed that contact interactions in AdS give rise to polynomial Mellin amplitudes, whose degree is related to the number of derivatives in the interaction vertex.
In fact, it is easy to see that contact interactions generate all possible polynomial 
Mellin amplitudes.
This proves the main result of \cite{JP,IdseJamie}, up to the intriguing possibility of non-local AdS interactions associated with analytic but non-polynomial Mellin
amplitudes.

There are several open questions worth studying in the future.
Firstly, it is natural to ask what is the Regge limit of the Mellin amplitudes.
The analogy with scattering amplitudes suggests that, for the
four point amplitude, it corresponds to
large $s_{12}$ with fixed $s_{13}$. However, it is not clear
that this controls the Regge limit of the four point function
as defined in \cite{ourEikonal,CornalbaRegge,ourBFKL,Mythesis}.
Secondly, it would be very interesting to generalize formula (\ref{mainformula}) for the flat space limit to the case of massive external particles in the scattering amplitude. In particular, this would allow
us to relate decay rates of excited string states in flat space
to three point functions non-BPS operators in SYM at large t'Hooft coupling.
Another important generalization, is to define Mellin amplitudes
for correlation functions of operators with spin. 
This should give a generalization of helicity for
conserved currents and tensors.
Finally,  the analogy with scattering amplitudes suggests that the 
Mellin amplitudes satisfy some unitarity bounds.
Perhaps, the analysis of \cite{Katz} can be useful in finding these
bounds.

\section*{Acknowledgements}
I wish to thank I. Heemskerk, J. Polchinski and J. Sully for collaboration
in the early stage of this work.  
I am also grateful for   discussions with M. Costa, T. Okuda and P. Vieira.
Research at the Perimeter Institute is supported in part by the Government
of Canada through NSERC and by the Province of Ontario through 
the Ministry of Research \& Innovation.
This research was supported in part by the 
National Science Foundation under
Grant No. NSFPHY05-51164.
This work was partially supported by the grant CERN/FP/109306/2009 and PTDC/FIS/099293/2008. Centro de F\'isica do Porto is partially funded by FCT through the POCI programme.

\appendix

\section{Mellin integration measure}
\label{appmeasure}

The precise definition of the integration measure $d\delta_{ij}$ in (\ref{genMellin})
was given in \cite{Mack,Symanzik}. Here, we quickly review it  for completeness.
Given a particular solution $\delta_{ij}^0$, with positive real part, of the constraints
(\ref{constraint}) we can write
\be
\delta_{ij}=\delta_{ij}^0+\sum_{k=1}^{\frac{1}{2}n(n-3)} c_{ij,k}\, s_k\ ,
\ee
where the real coefficients $c_{ij,k}=c_{ji,k}$ satisfy
\be
c_{ii,k}=0\ ,\ \ \ \ \ \ \ \ \ \ \ \ \ 
\sum_{j=1}^n c_{ij,k}=0\ .
\ee
We also demand that the $\left(\frac{1}{2}n(n-3)\right)^2$ coefficients $c_{ij,k}$ with
$2\le i<j\le n$, excepting $c_{23,k}$, which may be taken as the independent
ones, obey
\be
|\det c_{ij,k}| =1\ .
\ee
The integration measure is then given by
\be
\int d\delta_{ij} \ (\dots)=
 \int_{-i\infty}^{i\infty} \prod_{k=1}^{\frac{1}{2}n(n-3)} ds_k\ (\dots)\ .
\ee

\section{Harmonic analysis in hyperbolic space}

In the computation on Witten diagrams in Euclidean AdS$_{d+1}$
 it will be convenient to 
use a basis of harmonic functions in AdS. In this appendix we briefly summarize the necessary results.
For more details, we refer the reader to \cite{CornalbaRegge,ourBFKL,Mythesis}.
We choose   units where $R=1$.
An $SO(d+1,1)$ invariant function $F(X,Y)$ of two points in AdS can be expanded in a basis
of harmonic functions,
\be
F(X,Y)= \int_{-i\infty}^{i\infty} \frac{dc}{2\pi i} \hat{F}(c) \Omega_c(X,Y)\ ,
\ee
where 
\be
\Omega_c(X,Y)= N(c) 
\int_{\partial {\rm AdS}}  dP   \frac{1}{(-2 P\cdot X)^{h+c} (-2 P\cdot Y)^{h-c} }  \ ,  
\label{splitOmega}
\ee
with
\be
N(c)=\frac{\Gamma(h+c)\Gamma(h-c)}{2\pi^{2h }\Gamma(c)\Gamma(-c)} \ .
 \label{Nofc}
\ee
The function $\Omega_c$ is an even function of $c$ and satisfies
\be
\left( \nabla^2_X + h^2-c^2 \right) \Omega_c(X,Y)=0\ .
\ee
The transform $ \hat{F}(c)$ can be computed from
\be
\hat{F}(c)  = \frac{1}{\Omega_c(Y,Y)   } \int_{\rm AdS} dX\, \Omega_c(X,Y) F(X,Y)
\ee
where $\Omega_c(Y,Y) $ can be explicitly computed
 \be
\Omega_c(Y,Y) =  \frac{\pi^{h }\Gamma(h) }
{ \Gamma(2h) } N(c)\ .
\ee

\subsection{Bulk to bulk propagator}
\label{btobprop}

The bulk to bulk scalar propagator of dimension $\Delta$ is given by
\ba
G_{BB}(X,Y) &=& \frac{ \mathcal{C}_\Delta }{ u^{\Delta} } \ _2F_1\left(\Delta,\frac{2\Delta-d+1}{2},2\Delta-d+1,-\frac{4}{u} \right) \\
&=&\frac{1}{  (4\pi)^{h+1/2}}
 \int_{-i\infty}^{i\infty} \frac{dz}{2\pi i} \frac{\Gamma(z) \Gamma(\Delta-z) \Gamma(\frac{1}{2}-h+z)}{\Gamma(z+\Delta-2h+1)} \left(\frac{u}{4}\right)^{-z} \label{MBprop}
\ea
where
\be
u= (X-Y)^2 \
\ee
is the chordal distance in the embedding space $\mathbb{M}^{d+2}$.
When computing Witten diagrams, it will be convenient to use the harmonic space representation of the bulk to bulk propagator,
\be
G_{BB}(X,Y) =\int_{-i\infty}^{i\infty} \frac{dc}{2\pi i} \frac{1}{(\Delta-h)^2-c^2} \Omega_c(X,Y)\ .
\label{Splitprop}
\ee

We shall now check that (\ref{Splitprop}) is indeed equivalent to (\ref{MBprop}). 
This exercise will be useful to learn some basic techniques necessary to compute Witten diagrams. 
We start by writing 
\be
\Omega_c(X,Y)= \frac{1}{2\pi^{2h }\Gamma(c)\Gamma(-c)} 
\int_{\partial {\rm AdS}}  dP \int_0^\infty  \frac{dtd\bar{t}}{t\bar{t}}\,
 t^{h+c} \bar{t}^{h-c}  \,e^{2 t \,P\cdot X +2 \bar{t} \,P\cdot Y }  \ .
\ee
and performing the integral over $P$.
It is convenient to use Poincare coordinates $P=(P^+,P^-,P^\mu)=(1,x^2,x^\mu)$, with
lightcone coordinates for the $\mathbb{M}^{2}$ factor of $\mathbb{M}^{d+2}=\mathbb{M}^{2}\times\mathbb{R}^{d}$.
The vector $T=t X +\bar{t} Y  $ is future directed in $\mathbb{M}^{d+2}$. It is then convenient to pick coordinates where it is aligned with $(1,1,0)$. Then
\ba
 \int_{\partial {\rm AdS}}   dP  \,e^{2T\cdot P}=
 \int_{\mathbb{R}^d} dx e^{-|T|(1+x^2)} =
\frac{   \pi^h }{|T|^{h}}e^{-|T| }\ .
\ea
In the present case, we need
\ba
\int_0^\infty  \frac{dt d\bar{t} }{t \bar{t} }\,
 t ^{h+c } \bar{t} ^{h-c }  \,\frac{   \pi^h  }{|t X+\bar{t}Y|^{h}}e^{-|t X+\bar{t}Y|  }\ .
\ea
Inserting
\ba
1=\int_0^\infty ds \delta(s-t-\bar{t})
\ea
and scaling $t\to s t$ and $\bar{t}\to s \bar{t}$ we obtain
\ba
&&  \pi^h \int_0^\infty \frac{ds}{s}  \int_0^\infty  \frac{dt d\bar{t} }{t \bar{t} }\,
 t ^{h+c } \bar{t} ^{h-c }  \,\frac{   s^h   }{|t X+\bar{t}Y|^{h}}e^{-s |t X+\bar{t}Y|  } \delta(1-t-\bar{t})  \\
 &=&  \pi^h \int_0^\infty \frac{ds}{s}  \int_0^\infty  \frac{dt d\bar{t} }{t \bar{t} }\,
 t ^{h+c } \bar{t} ^{h-c }  \,    s^h  e^{s (t X+\bar{t}Y)^2 } \delta(1-t-\bar{t}) \ .
 \ea
After scaling $t \to t/\sqrt{s}$ and  $\bar{t}\to  \bar{t}/\sqrt{s}$ one can perform the integral over $s$ to obtain
\be
 2 \pi^h   \int_0^\infty  \frac{dt d\bar{t} }{t \bar{t} }\,
 t ^{h+c } \bar{t} ^{h-c }  \,       e^{ (tX+\bar{t}Y)^2  }   \label{IntBound}
 \ee
 This turns the expression for the bulk to bulk propagator into
\ba
G_{BB}(X,Y)
 &=&  2\pi^h \int_{-i\infty}^{i\infty} \frac{dc}{2\pi i }f(c)
  \int_0^\infty  \frac{dtd\bar{t}}{t\bar{t}}\,
 t^{h+c} \bar{t}^{h-c}  \,e^{ - (t+ \bar{t})^2    -u t \bar{t}}
\ea
We now use the representation
\be
e^{-u t \bar{t}}=\int \frac{dz}{2\pi i} \Gamma(z) (u t \bar{t})^{-z}
\ee
and perform the integrals over $t$ and $\bar{t}$
\be
\int_0^\infty  \frac{dtd\bar{t}}{t\bar{t}}\,
 t^{h-z+c} \bar{t}^{h-z-c}  \,e^{ - (t+ \bar{t})^2}=
 \frac{\Gamma (h-z) \Gamma
   \left(h-z+c\right)
   \Gamma
   \left(h-z-c\right)}{2
   \Gamma (2 h-2 z)}\ .
\ee
This gives
\be
G_{BB}(X,Y)
 =\frac{ 1}{2\pi^h  }\int \frac{dz}{2\pi i} \frac{\Gamma(z)\Gamma (h-z)  }{   \Gamma (2 h-2 z)}u^{-z}
 q(z)
\ee
where
\be
q(z)=\int_{-i\infty}^{i\infty} \frac{dc}{2\pi i }
  \frac{\Gamma  \left(h-z+c\right)   \Gamma   \left(h-z-c\right)}
  {\Gamma(c)\Gamma(-c)((\Delta-h)^2-c^2)}=\frac{\Gamma(\frac{1}{2}+h-z)\Gamma(\frac{1}{2}-h+z)\Gamma(\Delta-z)}{2\pi \Gamma(z+\Delta-2h+1)}\ .
\ee
To recover (\ref{MBprop}) one just needs to use the Legendre duplication formula of the $\Gamma$-function.

\section{Scalar exchange in AdS \label{appScalar}}

In this appendix, we compute the four point function associated to the Witten diagram of figure \ref{4ptexchangefig},
\be
A(P_i)=g^2 \int_{\rm AdS} dXdY G_{B\partial}(X,P_1)G_{B\partial}(X,P_3) G_{BB}(X,Y) G_{B\partial}(Y,P_2)G_{B\partial}(Y,P_4)\ . \label{ScalarAdSdef}
\ee
To compute the AdS integrals it is convenient to use the harmonic expansion (\ref{Splitprop}) of the bulk to bulk propagator. Reintroducing the factors of $R$, we have
\ba
G_{BB}(X,Y)&=&\frac{1}{R^{d-1} } \int_{-i\infty}^{i\infty} \frac{dc}{2\pi i } f(c)
\int_{\partial {\rm AdS}}  dP \int_0^\infty  \frac{dtd\bar{t}}{t\bar{t}}\,
 t^{h+c} \bar{t}^{h-c}  \,e^{2 t \,P\cdot X/R+2 \bar{t} \,P\cdot Y/R}  \ , \label{splitprop}
\ea
where
\be
f(c)=\frac{1}{2\pi^{2h }\Gamma(c)\Gamma(-c)}\frac{1}{(\Delta-h)^2-c^2}\ .
\ee
The correlation function (\ref{ScalarAdSdef}) can then be written as
\ba
A(P_i)&=&g^2 R^{5-d}\prod_{i=1}^4\frac{\mathcal{C}_{\Delta_i}}{\Gamma(\Delta_i)}
 \int_0^\infty \prod_{i=1}^4 \frac{dt_i}{t_i}t_i^{\Delta_i}
\int_{-i\infty}^{i\infty} \frac{dc}{2\pi i } f(c)
\int_0^\infty  \frac{dtd\bar{t}}{t\bar{t}}\,
 t^{h+c} \bar{t}^{h-c} \label{manyintegrals}\\
&&\int_{\partial {\rm AdS}}  dP  \int_{\rm AdS} d(X/R) e^{2 (t_1P_1+t_3P_3+t P)\cdot X/R}\int_{\rm AdS} d(Y/R) e^{2 (t_2P_2+t_4P_4+\bar{t} P)\cdot Y/R}\ .
\nonumber
\ea
The AdS integrals are of the form
\be
\int_{\rm AdS}  d(X/R) e^{2 Q\cdot X/R}
\ee
with $Q$ a future directed vector in $\mathbb{M}^{d+2}$.
Using Lorentz invariance, we can set $Q=|Q|(1,1,0)$ and $X=(X^+,X^-,X^\mu)=\frac{R}{z}(1,z^2+x^2,x^\mu)$,
with lightcone coordinates for the $\mathbb{M}^{2}$ factor of $\mathbb{M}^{d+2}=\mathbb{M}^{2}\times\mathbb{R}^{d}$.
Then
\ba
\int_{\rm AdS}  d(X/R) e^{2 Q\cdot X/R}&=& \int_0^\infty \frac{dz}{z} z^{-d} \int_{\mathbb{R}^d}dx e^{ -(1+z^2+x^2)|Q|/z}\\
&=& \pi^h \int_0^\infty \frac{dz}{z} (z|Q|)^{-h}  e^{ -(1+z^2 )|Q|/z}\\
&=& \pi^h \int_0^\infty \frac{dz}{z} z^{-h}  e^{ -z +Q^2/z}\ ,
\ea
and we can write the second line of (\ref{manyintegrals}) as follows
\be
\pi^{2h} \int_0^\infty \frac{dzd\bar{z}}{z\bar{z}} (z\bar{z})^{-h}  e^{ -z-\bar{z}}
\int_{\partial {\rm AdS}}  dP    e^{(t_1P_1+t_3P_3+t P)^2/z+ (t_2P_2+t_4P_4+\bar{t} P)^2/\bar{z}}\ .
\ee
The integral over $z$   can be easily done after scaling the variables $t_1$, $t_3$ and $t$ by $\sqrt{z}$.
Similarly for the integral over $\bar{z}$.
Thus, the correlation function now reads
\ba
A(P_i)&=&g^2 R^{5-d}\pi^{2h} \prod_{i=1}^4\frac{\mathcal{C}_{\Delta_i}}{\Gamma(\Delta_i)}
\int_0^\infty \prod_{i=1}^4 \frac{dt_i}{t_i}t_i^{\Delta_i} e^{-t_1 t_3 P_{13}- t_2t_4P_{24}}
\nonumber \\
&&
\int_{-i\infty}^{i\infty} \frac{dc}{2\pi i } f(c)
\Gamma\left(\frac{\Delta_1+\Delta_3+c-h}{2} \right)   \Gamma\left(\frac{\Delta_2+\Delta_4-c-h}{2} \right)
\\&&\int_0^\infty  \frac{dtd\bar{t}}{t\bar{t}}\,
 t^{h+c} \bar{t}^{h-c}
\int_{\partial {\rm AdS}}  dP    e^{2P\cdot \left(t (t_1P_1+ t_3P_3)+\bar{t}(t_2P_2+t_4P_4)\right)  } \ .
\nonumber
\ea
The integral  in the last line is exactly of the same form as the one we encountered in appendix \ref{btobprop}. It is given by
\be
2\pi^h \int_0^\infty  \frac{dtd\bar{t}}{t\bar{t}}\,
 t^{h+c} \bar{t}^{h-c}
    e^{  \left(t (t_1P_1+ t_3P_3)+\bar{t}(t_2P_2+t_4P_4)\right)^2  }\ ,
\ee
which gives
\ba
A(P_i)&=&g^2 R^{5-d}2\pi^{3h} \prod_{i=1}^4\frac{\mathcal{C}_{\Delta_i}}{\Gamma(\Delta_i)}
\int_{-i\infty}^{i\infty} \frac{dc}{2\pi i } f(c)
\nonumber \\
&&
\int_0^\infty  \frac{dtd\bar{t}}{t\bar{t}}\,
 t^{h+c} \bar{t}^{h-c}
\Gamma\left(   \frac{\Delta_1+\Delta_3+c-h}{2} \right)   \Gamma\left(\frac{\Delta_2+\Delta_4-c-h}{2} \right)
\\&&
\int_0^\infty \prod_{i=1}^4 \frac{dt_i}{t_i}t_i^{\Delta_i}
e^{-(1+t^2)t_1 t_3 P_{13}- (1+\bar{t}^2)t_2t_4P_{24}-t\bar{t}(t_1t_2P_{12}+t_1t_4P_{14}+t_2t_3P_{23}+t_3t_4P_{34})} \ .
\nonumber
\ea
Using the identity \cite{Symanzik}
\be
2\int_0^\infty \prod_{i=1}^n \frac{dt_i}{t_i}t_i^{\Delta_i} e^{-\sum_{i<j}^n t_i t_j Q_{ij}}=
\frac{1}{(2\pi i)^{n(n-3)/2}} \int_{\Sigma_n} d\delta_{ij} \prod_{i<j}^n \Gamma(\delta_{ij}) (Q_{ij})^{-\delta_{ij}} \ ,
\ee
with $Q_{ij}>0$,
we conclude that
\ba
M(\delta_{ij})&=&g^2 R^{5-d} 2\pi^{2h}
\int_{-i\infty}^{i\infty} \frac{dc}{2\pi i } f(c)\frac{
\Gamma\left(   \frac{\Delta_1+\Delta_3+c-h}{2} \right)   \Gamma\left(\frac{\Delta_2+\Delta_4-c-h}{2} \right)}{
\Gamma\left(\frac{\sum_i\Delta_i }{2}-h \right)}
\nonumber \\
&&
\int_0^\infty  \frac{dtd\bar{t}}{t\bar{t}}\,
 t^{h+c} \bar{t}^{h-c} (1+t^2)^{-\delta_{13}}(1+\bar{t}^2)^{-\delta_{24}}(t\bar{t})^{-\delta_{12}-\delta_{14}-\delta_{23}-\delta_{34}}\ .\\
\ea
After performing the integrals over $t$ and $\bar{t}$ one obtains formula (\ref{MBrep}).

\section{One-loop diagram in AdS}
\label{appLoop}

We would like to show that the Mellin amplitude associated to the 1-loop Witten diagram of figure 
\ref{1loopfig} is given by equation (\ref{MB1loop}).
This 1-loop diagram differs from the tree level diagram  (\ref{ScalarAdSdef}) computed in 
appendix \ref{appScalar} by the replacement
\be
G_{BB}(X,Y) \to G_{BB}(X,Y) G_{BB}(X,Y)\ ,
\ee
where the two propagators on the right need not have the same dimension.
Therefore, if we assume that 
\be
G_{BB}(X,Y) G_{BB}(X,Y)= \frac{1}{R^{2(d-1)}} \int_{-i\infty}^{i \infty} \frac{dc}{2\pi i} q(c) \Omega_c(X,Y)\ ,  \label{propsquared}
\ee
then all the computations of appendix   \ref{appScalar}  can be used with 
the simple substitution
\be
\frac{1}{(\Delta-h)^2-c^2} \to \frac{1}{R^{d-1}} q(c)
\ee
and equation (\ref{MB1loop}) follows.

We shall derive (\ref{propsquared}) as a particular case of a more general result. From now on we set $R=1$.
The problem is to find the harmonic decomposition  of the product
\be
 F_1(X,Y)  F_2(X,Y) = 
 \int_{-i\infty}^{i \infty} \frac{dc}{2\pi i} \hat{F}_{12}(c) \Omega_c(X,Y)\ ,
  \label{harmonicprod}
\ee
in terms of the harmonic expansion of each factor,
\be
 F_i(X,Y)  = 
 \int_{-i\infty}^{i \infty} \frac{dc}{2\pi i} \hat{F}_{i}(c) \Omega_c(X,Y)\ ,\ \ \ \ \ \ \ \ \ \ \ 
 i=1,2\ .
\ee
Inverting (\ref{harmonicprod}), we find
\ba
\hat{F}_{12}(c) &=& \frac{1}{\Omega_c(Y,Y)} \int_{\rm AdS} dX\, F_1(X,Y)  F_2(X,Y) \Omega_c(X,Y) \\
&=& \frac{1}{\Omega_c(Y,Y)} \int_{-i\infty}^{i \infty} \frac{dc_1dc_2}{(2\pi i)^2 } \hat{F}_{1}(c_1)\hat{F}_{2}(c_2) \Phi(c_1,c_2,c)
\ea
where
\be
\Phi(c_1,c_2,c_3)=
\int_{\rm AdS} dX\, \Omega_{c_1}(X,Y)  \Omega_{c_2}(X,Y) \Omega_{c_3}(X,Y)\ .
\ee

Using the split representation (\ref{splitOmega}) of the harmonic functions, we obtain
\be
\frac{\Phi(c_1,c_2,c_3)}{N(c_1)N(c_2)N(c_3)}=
\int_{\rm AdS} dX \int_{\partial{\rm AdS}} \prod_{i=1}^3\frac{dP_i}{
 (-2 P_i\cdot X)^{h+c_i} (-2 P_i\cdot Y)^{h-c_i}  }\ .
\ee
We start by performing the integral over $X$,
\ba
&&\int_{\rm AdS}dX  \prod_{i=1}^3\frac{1 }{
 (-2 P_i\cdot X)^{h+c_i}    } \\
  &=& \frac{\pi^h \Gamma\left(\frac{h+c_1+c_2+c_3}{2}\right)
 \Gamma\left(\frac{h+c_1+c_2-c_3}{2}\right)
  \Gamma\left(\frac{h+c_1+c_3-c_2}{2}\right)
   \Gamma\left(\frac{h+c_2+c_3-c_1}{2}\right)}{2 \Gamma(h+c_1)\Gamma(h+c_2)\Gamma(h+c_3) P_{12}^{\frac{h+c_1+c_2-c_3}{2}}P_{13}^{\frac{h+c_1+c_3-c_2}{2}}P_{23}^{\frac{h+c_2+c_3-c_1}{2}}}\ . \nonumber
\ea
This cubic AdS integral is well known \cite{Rastelli3pt,ourBFKL}.
We are then left with the following conformal integral
\be
I_{123}= \int_{\partial{\rm AdS}} \prod_{i=1}^3\frac{dP_i}{
 (-2 P_i\cdot Y)^{h-c_i} }
 \frac{1}{ P_{12}^{\frac{h+c_1+c_2-c_3}{2}}P_{13}^{\frac{h+c_1+c_3-c_2}{2}}P_{23}^{\frac{h+c_2+c_3-c_1}{2}}}\ .
\ee
As explained in appendix A of \cite{ourBFKL}, the strategy to evaluate this type of integrals always starts by introducing 
Schwinger parameters to exponentiate the denominators.
In the present case, we start by performing the integral over $P_3$. This gives
\be
I_{123}=\int_{\partial{\rm AdS}}  \frac{dP_1 dP_2}{ P_{12}^{2h}}  W(u)\ ,
\ee
where 
\ba
W(u)&=& \frac{\pi^h  \,u^{\frac{3h-c_1-c_2-c_3}{2}}}{
 \Gamma(h-c_3) \Gamma\left(\frac{h+c_1+c_3-c_2}{2}\right)
   \Gamma\left(\frac{h+c_2+c_3-c_1}{2}\right)}  
  \\
   &&  \int_0^\infty \frac{dt_1 dt_2 dt_3}{t_1 t_2 t_3}t_1^{\frac{h+c_1-c_2-c_3}{2}} t_2^{\frac{h+c_2-c_1-c_3}{2}} t_3^{c_3} \,e^{ -t_1-t_2-t_3 -u \frac{t_1 t_2}{t_3} }
\ea
 is a function of the unique invariant 
\be
u=\frac{P_{12}}{(-2 P_1\cdot Y)(-2 P_2\cdot Y)}\ ,
\ee
that can be formed using $P_1$, $P_2$ and $Y$.
Using 
\be
e^{-u \frac{t_1 t_2}{t_3}} =\int_{-i\infty}^{i \infty} \frac{dz}{2\pi i} \Gamma(-z) 
\left( u \frac{t_1 t_2}{t_3}\right)^z\ ,
\ee
we obtain
\be
I_{123}=\int_{-i\infty}^{i \infty} \frac{dz}{2\pi i} \frac{\pi^h \Gamma(-z) \Gamma(c_3-z)
 \Gamma\left(\frac{h+c_2-c_1-c_3}{2}+z\right) \Gamma\left(\frac{h+c_1-c_2-c_3}{2}+z\right) }{
 \Gamma(h-c_3) \Gamma\left(\frac{h+c_1+c_3-c_2}{2}\right)
   \Gamma\left(\frac{h+c_2+c_3-c_1}{2}\right)}  
I_{12} \ ,
\ee
where
\ba
I_{12}&=&\int_{\partial{\rm AdS}}  \frac{dP_1 dP_2}{ P_{12}^{2h}} 
  \left( \frac{P_{12}}{(-2 P_1\cdot Y)(-2 P_2\cdot Y)}\right)^{\frac{3h-c_1-c_2-c_3}{2} +z}\\
  &=& \frac{\pi^h \Gamma\left(\frac{h-c_1-c_2-c_3}{2}+z\right)}{
  \Gamma\left(\frac{3h-c_1-c_2-c_3}{2}+z\right)} 
  \int_{\partial{\rm AdS}}  \frac{dP_1}{ (-2P_i\cdot Y)^{2h}} \\
  &=& \frac{\pi^{2h} \Gamma(h)\Gamma\left(\frac{h-c_1-c_2-c_3}{2}+z\right)}{
  \Gamma(2h)\Gamma\left(\frac{3h-c_1-c_2-c_3}{2}+z\right)}   \ .
\ea
The integral over $z$ is precisely of the form of Barnes' second lemma,
\ba
&&\int \frac{dz}{2\pi i} \frac{  \Gamma(-z) \Gamma(c_3-z)
 \Gamma\left(\frac{h+c_2-c_1-c_3}{2}+z\right) \Gamma\left(\frac{h+c_1-c_2-c_3}{2}+z\right) 
 \Gamma\left(\frac{h-c_1-c_2-c_3}{2}+z\right)}{
\Gamma\left(\frac{3h-c_1-c_2-c_3}{2}+z\right)} \\
&=&\frac{  
 \Gamma\left(\frac{h+c_2-c_1-c_3}{2}\right) \Gamma\left(\frac{h+c_1-c_2-c_3}{2}\right) 
 \Gamma\left(\frac{h-c_1-c_2-c_3}{2}\right)
 \Gamma\left(\frac{h+c_2-c_1+c_3}{2}\right) \Gamma\left(\frac{h+c_1-c_2+c_3}{2}\right) 
 \Gamma\left(\frac{h-c_1-c_2+c_3}{2}\right) 
 }{\Gamma(h)\Gamma(h-c_1)\Gamma(h-c_2)} \ . \nonumber
 \ea
This gives
\be
I_{123}=
\frac{  \pi^{3h}
 \Gamma\left(\frac{h+c_2-c_1-c_3}{2}\right) \Gamma\left(\frac{h+c_1-c_2-c_3}{2}\right) 
 \Gamma\left(\frac{h-c_1-c_2-c_3}{2}\right) 
 \Gamma\left(\frac{h-c_1-c_2+c_3}{2}\right) 
 }{\Gamma(2h)\Gamma(h-c_1)\Gamma(h-c_2)\Gamma(h-c_3)}
\ee
and
\be
\frac{\Phi(c_1,c_2,c_3)}{N(c_1)N(c_2)N(c_3)}= \frac{\pi^{4h}}{2\Gamma(2h)}
\frac{ \prod_{\{\sigma_i=\pm\}}  \Gamma\left(\frac{h +\sigma_1c_1+\sigma_2c_2+\sigma_3c_3}{2}\right)  }{\prod_{i=1}^3 \Gamma(h+c_i)\Gamma(h-c_i)}\ .
\ee
Using (\ref{Nofc}) we obtain
\be
\Phi(c_1,c_2,c_3)= \frac{1}{16\pi^{2h}\Gamma(2h)} \Theta(c_1,c_2,c_3)\ ,
\ee
with $\Theta(c_1,c_2,c_3)$ given by equation (\ref{Theta}).
Finally, we conclude that
\be
\hat{F}_{12}(c) = \frac{\Gamma(c)\Gamma(-c)}{8\pi^h\Gamma(h) \Gamma(h+c)\Gamma(h-c)}
 \int_{-i\infty}^{i\infty} \frac{dc_1dc_2}{(2\pi i)^2 } \hat{F}_{1}(c_1)\hat{F}_{2}(c_2) \Theta(c_1,c_2,c)\ .
\ee

\section{Angular integral \label{angularint}}

\begin{figure} 
\begin{center}$
\begin{array}{cc}
\includegraphics[height=5cm]{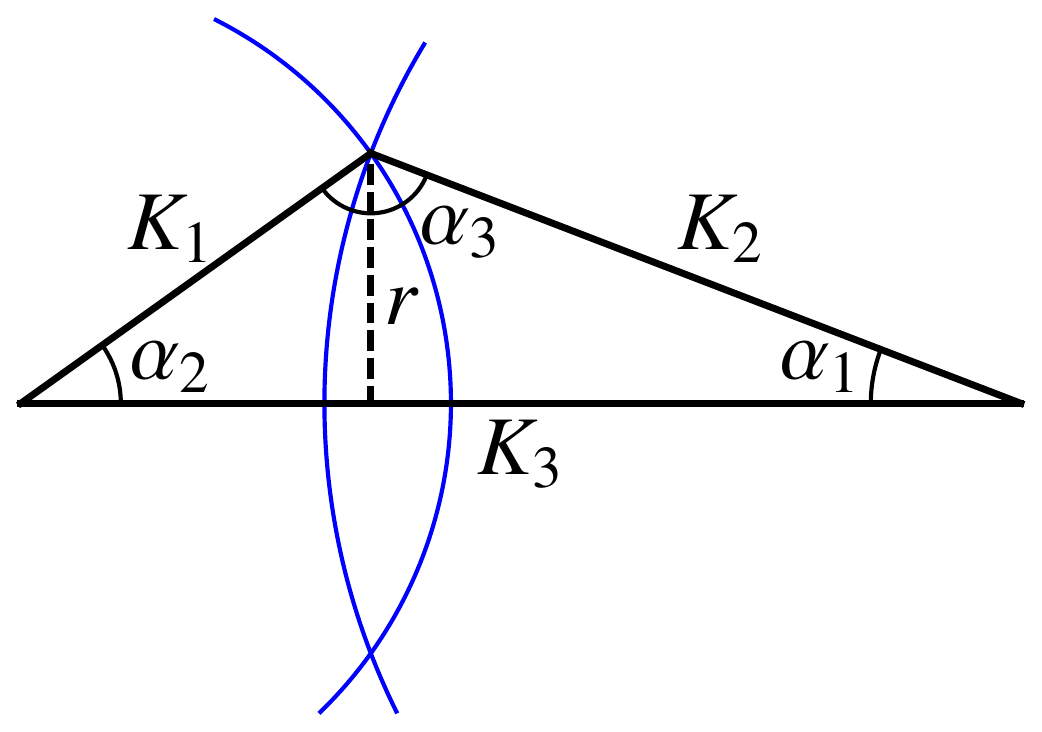} &
\includegraphics[height=7cm]{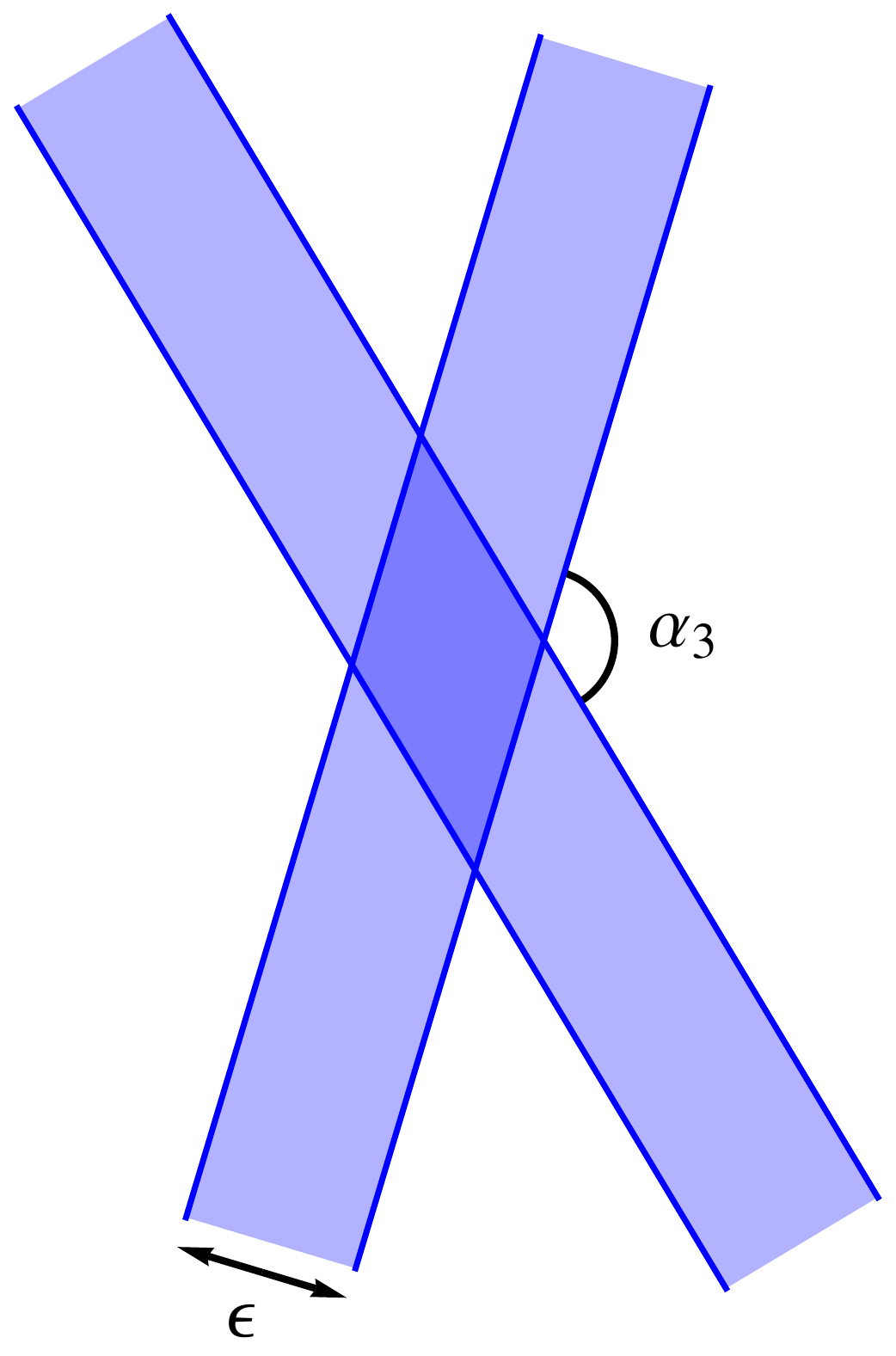}
\end{array}$
\end{center}
\caption{\small On the left, we sketch the intersection of two $d$-spheres of radius $K_1$ and $K_2$ and whose centres are separated by a distance $K_3$. The radius of the resulting $(d-1)$-sphere is $r$. On the right, we zoom into the upper vertex of the triangle on the left and give the spheres a finite thickness $\epsilon$. It is easy to see that the area of the intersection is given by $\epsilon^2/\sin\alpha_3$.
}
\label{triangle}
\end{figure}

The goal of this appendix is to compute the integral
\be
I(K_1,K_2,K_3)= \int_{S^{d}} d\hat{K}_1d\hat{K}_2  \delta^{d+1}(K_1\hat{K}_1 +K_2\hat{K}_2+K_3\hat{K}_3)\ ,
\ee
where $\hat{K}_i \in S^d$ and $K_i >0$. 
The integral $I(K_1,K_2,K_3)$ is  invariant under permutations of its arguments.
This is obvious from 
\be
 \int_{S^{d}} d\hat{K}_1d\hat{K}_2  d\hat{K}_3 \delta^{d+1}(K_1\hat{K}_1 +K_2\hat{K}_2+K_3\hat{K}_3)  = V_{S^d}\ I(K_1,K_2,K_3)\ ,
\ee
where $V_{S^d}$ is the volume of the $d$-dimensional sphere.
Another way to write our integral is
\ba
I(K_1,K_2,K_3)&=&    \int_{\mathbb{R}^{d+1}} dKdK' 
\frac{\delta(|K|-K_1)\delta(|K'|-K_2)
 \delta^{d+1}(K +K'+K_3\hat{K}_3)}{
 (K_1K_2)^d}\\
 &=&
 \frac{1}{(K_1K_2)^d} \int_{\mathbb{R}^{d+1}} dK \delta(|K|-K_1)\delta(|K+K_3\hat{K}_3|-K_2)\ . 
 \label{2deltas}
 \ea
 The first delta-function in (\ref{2deltas}) says that $K$ lays on the $d$-sphere of radius $K_1$ centred at the origin
 and the second delta-function says it belongs to the $d$-sphere of radius $K_2$ centred at the point
 $-K_3\hat{K}_3$ (see figure \ref{triangle}). 
 It is then clear that $I(K_1,K_2,K_3)$ vanishes if it is not possible to form
 a triangle with sides $K_1$, $K_2$ and $K_3$.
 Naively, the answer would be given by the volume of the $(d-1)$-sphere defined by 
 the intersection of the two $d$-spheres,
 \be
 \int_{\mathbb{R}^{d+1}} dK \delta(|K|-K_1)\delta(|K+K_3\hat{K}_3|-K_2) \to V_{S^{d-1}} r^{d-1}
 \ee
 where $r$ is the radius of the $(d-1)$-sphere as shown in figure \ref{triangle} and is given by
 \be
 r=\frac{2 \,Area(K_1,K_2,K_3)}{K_3}\ ,
 \ee
 with 
 \be
 Area(K_1,K_2,K_3) = \frac{1}{4}\sqrt{ 2K_1^2K_2^2+ 2K_1^2K_3^2+2K_2^2K_3^2 -K_1^4-K_2^4-K_3^4}
 \ee
 being the area of the triangle formed by $K_1$, $K_2$ and $K_3$.
 This gives
 \be
 I(K_1,K_2,K_3) \to V_{S^{d-1}}  \frac{\left(2\, Area(K_1,K_2,K_3)\right)^{d-1}} {(K_1K_2K_3)^d}K_3 \ ,
 \ee
 which can not be correct because it does not respect the full permutation symmetry of $I$.
In fact, we were not careful about the induced measure on the $(d-1)$-sphere. To see the problem,
consider a regulated version of the delta-functions in (\ref{2deltas}),
\be
\delta(t) \to \delta_\epsilon (t) =\left\{
\begin{array}{ll}
1/\epsilon  \ , & |t|\le \epsilon/2 \\
0\ ,\ \ \ \ \ & |t|>\epsilon/2
 \end{array} \right.
 \ee
The integral in (\ref{2deltas}) is then given by the $(d+1)$-dimensional volume of the intersection of
two thin $d$-spheres with thickness $\epsilon$, divided by $\epsilon^2$. As depicted in figure \ref{triangle},
the result is not simply the volume of the $(d-1)$-sphere because, in general, the two $d$-spheres do not intersect perpendicularly. However, this effect is very easy to take into account and it only gives an extra factor of $1/\sin\alpha_3$ (see figure \ref{triangle}).
The right answer is then
 \ba
 I(K_1,K_2,K_3) &=& V_{S^{d-1}}  \frac{\left(2\, Area(K_1,K_2,K_3)\right)^{d-1}} {(K_1K_2K_3)^d}
 \frac{K_3}{\sin \alpha_3} \\
&=&\frac{2\pi^{\frac{d}{2}}}{\Gamma\left(\frac{d}{2} \right)}
 \frac{\left( 2\,Area(K_1,K_2,K_3) \right)^{d-2}}{(K_1K_2K_3)^{d-1}}\ .
\ea

\section{Primary double-trace operators}
\label{DTapp}

The conformal algebra is \cite{CFTbook}
\begin{align}
[D ,P_\mu]&=iP_\mu \ , \ \ \ \ \ \ \  
[D ,K_\mu] =-iK_\mu \ ,\ \ \ \ \ \ \ \  
[K_\mu,P_\nu]  =2i(\eta_{\mu\nu} D  -L_{\mu\nu}) \ ,\nonumber \\
[K_\rho,L_{\mu\nu}]&=i(\eta_{\rho\mu} K_\nu -\eta_{\rho\nu}K_\mu) \ , \ \ \ \ \ \ \ \ \ \
 \ \ \  
[ P_{\rho},L_{\mu\nu} ]  = i (\eta_{\rho\mu} P_{\nu} -\eta_{\rho\nu} P_{\mu})\ ,  
\\
[ L_{\mu\nu},L_{\rho\sigma}]&=i(\eta_{\nu\rho}L_{\mu\sigma}  
+\eta_{\mu\sigma}L_{\nu\rho} -\eta_{\mu\rho}L_{\nu\sigma}
  -\eta_{\nu\sigma}  L_{\mu\rho}\ .\nonumber 
) 
\end{align}
A primary operator $\Ocal$ of dimension $\Delta$ is defined by
\be
D\, \Ocal = i \Delta\, \Ocal\ , \ \ \ \ \ \ \ \ \ \ \ 
K_\mu\, \Ocal =0\ .
\ee
We now wish to construct new primaries by taking the normal ordered product
of descendants of two primaries $\Ocal_1$ and $\Ocal_2$. 
At dimension $\Delta_1+\Delta_2+k$ we have a large number of possible operators.
For example, at $k=2$ we have
\be
P_{\mu}P_{\nu}\Ocal_1\,\Ocal_2\ ,\ \ \ \ \ \ \ \ 
P_{\mu}\Ocal_1\,P_{\nu}\Ocal_2\ ,\ \ \ \ \ \ \ \ 
\Ocal_1\,P_{\mu}P_{\nu}\Ocal_2\ .
\ee
The dimension $N(k)$ of this vector space at level $k$ is
\be
N(k)=\sum_{m=0}^k \frac{(m+d-1)!}{m!(d-1)!}\frac{(k-m+d-1)!}{(k-m)!(d-1)!}\ .
\label{levelkdim}
\ee
This vector space can be decomposed into primary operators and descendants of
primaries with lower $k$. For $k=2$ we have
\be
P_{\mu}P_{\nu} \left[\Ocal_1\Ocal_2\right]^{( 0)}\ ,\ \ \ \ \ \ \ \ 
P_{\mu} \left[\Ocal_1\Ocal_2\right]^{( 1)}_{\nu}\ ,\ \ \ \ \ \ \ \
  \left[\Ocal_1\Ocal_2\right]^{( 2)}_{\mu\nu}\ .
\ee 
where $ \left[\Ocal_1\Ocal_2\right]^{(k)}_{\mu_1\dots\mu_k}$ denotes a primary at level $k$.
At general level $k$ we find the decomposition
\be
N(k)=\sum_{m=0}^k \frac{(m+d-1)!}{m!(d-1)!}N_p(k-m)\ , \label{primarysum}
\ee
where $N_p(k)$ is the dimension of the vector space of primaries at level $k$.
Comparing (\ref{levelkdim}) with (\ref{primarysum}) we conclude that
\be
N_p(k)=\frac{(k+d-1)!}{k!(d-1)!}\ .
\ee
This is precisely the number of components of a symmetric tensor with $k$ indices.
We can further split this tensor into irreducible representations of the rotation group
$SO(d)$ (basically by removing traces).
We conclude that the primary double-trace operators are labeled by the spin $l\ge0$ (totally symmetric and traceless tensor with $l$ indices) and the dimension 
$\Delta_1+\Delta_2+2n+l$, where $n\ge0$ is directly related to the number of traces.
Our counting argument shows that there is only one primary for each label $(n,l)$.

The explicit form of these primary operators is the following 
\be
V^{\alpha_1 \dots  \alpha_{l} } 
 \left[\Ocal_1\Ocal_2\right]^{(2n+l)}_{\alpha_1 \dots  \alpha_{l}} 
 = \sum_{k_1,k_2,u_1,u_2,m \ge0 \atop k_1+k_2=l ,\,  u_1+u_2+m=n} 
T(k_1,k_2,u_1,u_2,m) \,
a(k_1,k_2,u_1,u_2,m) \ ,
\label{doubletrace}
\ee
where $T(k,l-k,u_1,u_2,m) $ is given by
\be 
  V^{\alpha_1 \dots  \alpha_{l} }\ 
P_{\alpha_1}\dots P_{\alpha_{k } } P_{\mu_1}\dots P_{\mu_m} (P^2)^{u_1}\Ocal_1\ 
P_{\alpha_{k +1}}\dots P_{\alpha_{l} } 
P^{\mu_1}\dots P^{\mu_m} (P^2)^{u_2} \Ocal_2\ ,
\ee
with the tensor $V^{\alpha_1 \dots  \alpha_{l} }$ traceless and symmetric.
The conformal dimension of this operator is easily found from the commutation relations,
\be
D \left[\Ocal_1\Ocal_2\right]^{(2n+l)}_{\alpha_1 \dots  \alpha_{l}} =i(\Delta_1+\Delta_2+2n+l) \left[\Ocal_1\Ocal_2\right]^{(2n+l)}_{\alpha_1 \dots  \alpha_{l}}\ ,
\ee
independently of the coefficients $a(k_1,k_2,u_1,u_2,m) $.
The condition 
\be
K_\mu \left[\Ocal_1\Ocal_2\right]^{(2n+l)}_{\alpha_1 \dots  \alpha_{l}}=0
\ee
determines the coefficients
$a(k_1,k_2,u_1,u_2,m)$. 
Finding the solution to the general case is a non-trivial task.
 However, in the minimal twist case ($n=0$) the equations simplify and we can write the coefficients $b(k)\equiv a(k ,l-k,0,0,0)$ in closed form.
First consider the action of $K_{\mu}$
on a descendent,
\begin{align}
 K_{\mu}
P_{\alpha_1}\dots P_{\alpha_{k } }  \Ocal_1 &=
2\sum_{s>r\ge1}^k \eta_{\alpha_{r}\alpha_{s}} 
P_\mu P_{\alpha_1}\dots \hat{P}_{\alpha_{r}}\dots \hat{P}_{\alpha_{s}}\dots P_{\alpha_{k } }  \Ocal_1\\
&-2(\Delta_1+k-1)\sum_{r=1}^k \eta_{\mu\alpha_r} 
P_{\alpha_1}\dots \hat{P}_{\alpha_{r}}\dots  P_{\alpha_{k } }  \Ocal_1\ ,
\nonumber
\end{align}
where $\hat{P}_{\nu}$ denotes that $P_{\nu}$ does not appear in the list.
Using this result, it is easy to see that
\begin{align}
& V^{\alpha_1 \dots  \alpha_{l} } 
 K_\mu \left[\Ocal_1\Ocal_2\right]^{(l)}_{\alpha_1 \dots  \alpha_{l}} =
 -2 \, V_\mu^{\ \alpha_2 \dots \alpha_l} \\& \sum_{k=1}^l  
    \Big(
k(\Delta_1+k-1)  b(k) + (l-k+1)(\Delta_1+l-k)  b(k-1) 
  \Big)
  P_{\alpha_2}\dots P_{\alpha_k} \Ocal_1\, 
  P_{\alpha_{k+1}}\dots P_{\alpha_l} \Ocal_2 \ .\nonumber
\end{align}
Setting this to zero provides a recursion relation for the coefficients $b(k)$.
The unique solution, up to normalization, is
\be
b(k)=\frac{(-1)^k}{
\Gamma(k+1) \Gamma(l-k+1)\Gamma(\Delta_1+k)\Gamma(\Delta_2+l-k) 
 }\ .
\ee
This generalizes the result of \cite{Mikhailov:2002bp} which analyzed the case
 when $\Ocal_1=\Ocal_2$ is a massless free scalar field.

\bibliographystyle{utphys}
\bibliography{mybib}
 
\end{document}